\documentclass[11pt,english]{article}
\usepackage[T1]{fontenc}
\usepackage[latin9]{inputenc}
\usepackage{geometry}
\usepackage{array}
\usepackage{float}
\usepackage{textcomp}
\usepackage{multirow}
\usepackage{amsmath}
\usepackage{amssymb}
\usepackage{graphicx}
\usepackage{setspace}
\usepackage[authoryear]{natbib}
\makeatletter

\providecommand{\tabularnewline}{\\}

\@ifundefined{date}{}{\date{}}
\usepackage{natbib}
\usepackage{graphicx, graphics, color, amsfonts}
\usepackage{hyperref}
\usepackage[T1]{fontenc}
\usepackage{rotating}
\usepackage{amsmath}
\hypersetup{colorlinks=true}
\hypersetup{linkcolor=black}
\hypersetup{urlcolor=black}
\hypersetup{citecolor=black}
\usepackage{lscape}
\usepackage{booktabs,fixltx2e}
\usepackage{threeparttable}
\floatstyle{plaintop}
\restylefloat{table}
\usepackage{subfig}
\usepackage{xfrac}
\usepackage{cancel}
\usepackage[font=footnotesize, tableposition=bottom]{caption}
\usepackage{setspace}
\usepackage{algorithm}
\usepackage{pdflscape}
\usepackage{afterpage}
\usepackage{xfrac}
\usepackage{cancel}
\usepackage{amsmath}
\usepackage[font=footnotesize]{caption}
\usepackage[symbol]{footmisc}
\usepackage{setspace}

\makeatother

\usepackage{babel}
\begin{document}
\title{Bayesian Optimization of Hyperparameters from Noisy Marginal Likelihood
Estimates}
\author{Oskar Gustafsson\emph{\footnotesize{}$^{1}$}\thanks{Corresponding author: Oskar Gustafsson, Department of Statistics SE-10691
Stockholm, Sweden. Email:oskar.gustafsson@stat.su.se. Phone:(+46)739692774}, Mattias Villani\emph{\footnotesize{}$^{1,2}$} and Pär Stockhammar\emph{\footnotesize{}$^{1,3}$}}

\maketitle
\vspace{-0.9cm}
\begin{center}
\emph{\footnotesize{}$^{1}$Department of Statistics, Stockholm University}\\
\emph{\footnotesize{}$^{2}$Department of Computer and Information
Science, Linköping University}\\
\emph{\footnotesize{}$^{3}$Sveriges Riksbank}{\footnotesize\par}
\par\end{center}
\begin{abstract}
Bayesian models often involve a small set of hyperparameters determined
by maximizing the marginal likelihood. Bayesian optimization is a
popular iterative method where a Gaussian process posterior of the
underlying function is sequentially updated by new function evaluations.
An acquisition strategy uses this posterior distribution to decide
where to place the next function evaluation. We propose a novel Bayesian
optimization framework for situations where the user controls the
computational effort, and therefore the precision of the function
evaluations. This is a common situation in econometrics where the
marginal likelihood is often computed by Markov chain Monte Carlo
(MCMC) or importance sampling methods, with the precision of the marginal
likelihood estimator determined by the number of samples. The new
acquisition strategy gives the optimizer the option to explore the
function with cheap noisy evaluations and therefore find the optimum
faster. The method is applied to estimating the prior hyperparameters
in two popular models on US macroeconomic time series data: the steady-state
Bayesian vector autoregressive (BVAR) and the time-varying parameter
BVAR with stochastic volatility. The proposed method is shown to find
the optimum much quicker than traditional Bayesian optimization or
grid search.
\end{abstract}

\section{Introduction}

The trend in econometrics is to use increasingly more flexible models
that give a richer description of the economy, particularly for prediction
purposes. As the model complexity increases, the estimation problems
get more involved, and computationally costly MCMC methods are often
used to sample from the posterior distribution.

Most models involve a relatively small set of hyperparameters that
needs to be chosen by the user. For example, consider the steady-state
BVAR model \citep{villani2009}, which is widely used among practitioners
and professional forecasters \citep{karlsson2013}, and used in Section
\ref{sec:Empirical-applications} for illustration. The choice of
the prior distribution in BVARs is often reduced to the selection
of a small set of prior hyperparameters. Some of these hyperparameters
can be specified subjectively by experts, for example, the steady-state
is usually given a rather informative subjective prior. Other prior
hyperparameters control the smoothness/shrinkage properties of the
model and are less easy to specify subjectively.

\citet{giannone2015} proposed to treat these hard-to-specify prior
hyperparameters as unknown parameters and explore the joint posterior
of the hyperparameters, the VAR dynamics, and the shock covariance
matrix. This is a statistically elegant approach which works well
when the marginal likelihood is available in closed form and is easily
evaluated. However, the marginal likelihood is rarely available in
closed form. The BVARs with conjugate priors considered in \citet{carriero2012},
and \citet{giannone2015} are an exception, but already the steady-state
VAR needs MCMC methods to evaluate the marginal likelihood. It is
of course always an option to sample the hyperparameters jointly with
the other model parameters using Metropolis-Hastings (MH) or Hamiltonian
Monte Carlo (HMC), but this likely leads to inefficient samplers since
the parameter spaces are high-dimensional and the posterior of the
hyperparameters are often quite complex, see e.g. the application
in Section \ref{sec:Time-varying-parameter-BVAR}.

Most practitioners also seem to prefer to fix the hyperparameters
before estimating the other model parameters. \citet{carriero2012}
propose a brute force optimization approach where the marginal likelihood
is evaluated over a grid. This is computationally demanding, especially
if a simulation based method has to be used for computing the marginal
likelihood. Since the marginal likelihood in \citet{giannone2015}
is available in closed form, one can readily optimize it using a standard
gradient based optimizer with automatic differentiation, but this
is again restricted to models with conjugate priors. The vast majority
of applications instead use so-called conventional values for the
hyperparameters, dating back to \citet{doan1984forecasting}, which
were found to be optimal on a specific historical dataset but are
likely to be suboptimal for other datasets. Hence, there is a real
need for a fast method for optimizing the marginal likelihood over
a set of hyperparameters when every evaluation of the marginal likelihood
is a noisy estimate from a computationally costly full MCMC run.

Bayesian optimization (BO) is an iterative optimization technique
originating from machine learning. BO is particularly suitable for
optimization of costly noisy functions in small to moderate dimensional
parameter spaces (\citealp{brochu2010} and \citealp{snoek2012practical})
and is therefore well suited for marginal likelihood optimization.
The method treats the underlying objective function as an unknown
object that can be inferred by Bayesian inference by evaluating the
function at a finite number of points. A Gaussian process prior expresses
Bayesian prior beliefs about the underlying function, often just containing
the information that the function is believed to have a certain smoothness.
Bayes theorem is then used to sequentially update the Gaussian process
posterior after each new function evaluation. Bayesian optimization
uses the most recently updated posterior of the function to decide
where to optimally place the next function evaluation. This so-called
\emph{acquisition strategy} is a trade-off between: i) \emph{exploiting}
the available knowledge about the function to improve the current
maxima and ii) \emph{exploring} the function to reduce the posterior
uncertainty about the objective function.

Our paper proposes a novel framework for Bayesian optimization when
the user can control the precision and computational cost of each
function evaluation. The framework is quite general, but we focus
mainly on the situation when the noisy objective function is a marginal
likelihood computed by MCMC. This is a very common situation in econometrics
using, for example, the estimators in \citet{chib1995marginal}, \citet{chib2001marginal},
and \citet{geweke1999using}. The precision of the marginal likelihood
estimate at each evaluation point is then implicitly chosen by the
user via the number of MCMC iterations. This makes it possible to
use occasional cheap noisy evaluations of the marginal likelihood
to quickly explore the marginal likelihood over hyperparameter space
during the optimization. Our proposed acquisition strategy can be
seen as jointly deciding where to place the new evaluation but also
how much computational effort to spend in obtaining the estimate.
We implement this strategy by a stopping rule for the MCMC sampling
combined with an auxiliary prediction model for the computational
effort at any new evaluation point; the auxiliary prediction model
is learned during the course of the optimization.

We apply the method to the steady-state BVAR \citep{villani2009}
and the time-varying parameter BVAR with stochastic volatility \citep{chan2018}
and demonstrate that the new acquisition strategy finds the optimal
hyperparameters faster than traditionally used acquisition functions.
It is also substantially faster than a grid search and finds a better
optimum.

The outline of the paper is a follows. Section \ref{sec:Hyperparameter-estimation-by}
introduces the problem of inferring hyperparameters from an estimated
marginal likelihood. Section \ref{sec:Bayesian-optimization-for}
gives the necessary background on Gaussian processes and Bayesian
optimization and introduces our new Bayesian optimization framework.
Section \ref{sec:Simulation-experiment} illustrates and evaluates
the proposed algorithm in a simulation study. Sections \ref{sec:Empirical-applications}
and \ref{sec:Time-varying-parameter-BVAR} assess the performance
of the proposed algorithm in empirical examples, and the final section
concludes.

\section{Hyperparameter estimation from an estimated marginal likelihood\label{sec:Hyperparameter-estimation-by}}

To introduce the problem of using an estimated marginal likelihood
for learning a model's hyperparameters we consider the selection of
hyperparameters in the popular class of Bayesian vector autoregressive
models (BVARs) as our running example.

\subsection{Hyperparameter estimation}

Consider the standard BVAR model
\begin{equation}
\boldsymbol{y}_{t}=\sum_{k=1}^{K}\boldsymbol{\Pi}_{k}\boldsymbol{y}_{t-k}+\boldsymbol{\varepsilon}_{t},
\end{equation}
where $\left\{ \boldsymbol{\varepsilon}_{t}\right\} _{t=1}^{T}$ are
iid $N(\boldsymbol{0},\boldsymbol{\Sigma})$. A simplified version
of the Minnesota prior (see e.g. \citet{karlsson2013}) without cross-equation
shrinkage is of the form

\begin{equation}
(\boldsymbol{\Pi},\boldsymbol{\Sigma})\sim MNIW(\underline{\boldsymbol{\Pi}},\underline{\boldsymbol{\Omega}}_{\Pi},\underline{\boldsymbol{S}},\underline{\nu}),
\end{equation}
with

\begin{equation}
\boldsymbol{\Sigma}\sim IW(\underline{\boldsymbol{S}},\underline{\nu}),\:\text{vec}(\boldsymbol{\Pi}')|\boldsymbol{\Sigma}\sim N(\text{vec}(\boldsymbol{\Pi}'),\boldsymbol{\Sigma}\otimes\underline{\boldsymbol{\Omega}}_{\Pi}).
\end{equation}
where $\text{vec}(\underline{\boldsymbol{\Pi}})$ and $\underline{\boldsymbol{\Omega}}_{\Pi}$
denotes the prior mean and covariance of the coefficient matrix, $\underline{\boldsymbol{S}}$
is the prior scale matrix with the prior degrees of freedom, $\underline{\nu}$.
The diagonal elements of $\underline{\boldsymbol{\Omega}}_{\Pi}$
are given by

\begin{equation}
\underline{\omega}_{ii}=\frac{\lambda_{1}^{2}}{(l^{\lambda_{3}}s_{r})^{2}},\text{ for lag }l\text{ of variable }r,\,i=(l-1)p+r,
\end{equation}
where $\lambda_{1}$ controls the \emph{overall shrinkage} and $\lambda_{3}$
the \emph{lag-decay shrinkage} set by the user, $s_{r}$ denotes the
estimated standard deviation of variable $r$. The fact that we do
not use the additional cross-equation shrinkage hyperparameter, $\lambda_{2}$,
makes this prior conjugate to the VAR likelihood, a fact that will
be important in the following. It has been common practice to use
standard values that dates back to \citet{doan1984forecasting}, but
there has been a renewed interest to find values that are optimal
for the given application (see e.g. \citet{banbura}, \citet{carriero2012}
and \citet{giannone2015}). Two main approaches have been proposed.
First, \citet{giannone2015} proposed to sample from the joint posterior
distribution using the decomposition
\begin{equation}
p(\boldsymbol{\beta},\boldsymbol{\theta}\vert\mathbf{y}_{1:T})=p(\boldsymbol{\beta}\vert\boldsymbol{\theta},\mathbf{y}_{1:T})p(\boldsymbol{\theta}\vert\mathbf{y}_{1:T}),
\end{equation}
where $\boldsymbol{\beta}=(\Pi,\Sigma)$ and $p(\boldsymbol{\theta}\vert\mathbf{y}_{1:T})$
is the marginal posterior distribution of the hyperparameters. The
algorithm samples from $p(\boldsymbol{\theta}\vert\mathbf{y}_{1:T})$
using Metropolis-Hastings (MH) and then samples directly from $p(\boldsymbol{\beta}\vert\boldsymbol{\theta},\mathbf{y}_{1:T})$
for each $\boldsymbol{\theta}$ draw by drawing $\Pi$ and $\Sigma$
from the Normal-Inverse Wishart distribution. There are some limitations
to using this approach. First, the $p(\boldsymbol{\theta}\vert\mathbf{y}_{1:T})$
can be multimodal (see e.g. the application in Section \ref{sec:Time-varying-parameter-BVAR})
and it can be hard to find a good MH proposal density, making the
sampling time-consuming. Second, practitioners tend to view hyperparameter
selection as similar to model selection and want to determine a fixed
value for $\boldsymbol{\theta}$ once and for all early in the model
building process.

\citet{carriero2012} propose an exhaustive grid search to find the
$\boldsymbol{\theta}$ that maximizes $p(\boldsymbol{\theta}\vert\mathbf{y}_{1:T})$
and then uses that optimal $\boldsymbol{\theta}$ throughout the remaining
analysis. The obvious drawback here is that a grid search is very
costly, especially if we have non-conjugate priors and more than a
couple of hyperparameters.

A problem with both the approach in \citet{giannone2015} and \citet{carriero2012}
is that for most interesting models $p(\boldsymbol{\theta}\vert\mathbf{y}_{1:T})$
is not available in closed form. Even the Minnesota prior with cross-equation
shrinkage is no longer a conjugate prior and $p(\boldsymbol{\theta}\vert\mathbf{y}_{1:T})$
is intractable. In fact, most Bayesian models used in practice have
intractable $p(\boldsymbol{\theta}\vert\mathbf{y}_{1:T})$, including
the steady-state BVAR \citep{villani2009} and the TVP-SV BVAR \citep{primiceri2005}
used in Section \ref{sec:Empirical-applications} and \ref{sec:Time-varying-parameter-BVAR}.

When $p(\boldsymbol{\theta}\vert\mathbf{y}_{1:T})$ is intractable,
MCMC or other simulation based methods like Sequential Monte Carlo
(SMC) are typically used to obtain a noisy estimate of $p(\boldsymbol{\theta}\vert\mathbf{y}_{1:T})$.
Since 
\begin{equation}
p(\boldsymbol{\theta}\vert\mathbf{y}_{1:T})\propto p(\mathbf{y}_{1:T}\vert\boldsymbol{\theta})p(\boldsymbol{\theta}),
\end{equation}
where $p(\mathbf{y}_{1:T}\vert\boldsymbol{\theta})=\int p(\mathbf{y}_{1:T}\vert\boldsymbol{\theta},\beta)p(\beta\vert\boldsymbol{\theta})d\beta$
is the marginal likelihood, this problem goes under the heading of
(log) marginal likelihood estimation. We will therefore frame our
method as maximizing the log marginal likelihood; maximization of
$p(\boldsymbol{\theta}\vert\mathbf{y}_{1:T})$ is achieved by simply
adding the log prior, $\log p(\boldsymbol{\theta})$, to the objective
function.

\citet{chib1995marginal} proposes an accurate way of computing a\emph{
simulation-consistent} estimate of the marginal likelihood when the
posterior can be obtained via Gibbs sampling, which is the case for
many econometric models. We refer to \citet{chib1995marginal} for
details about the marginal likelihood estimator and its approximate
standard error.

Estimating the marginal likelihood for the TVP-SV BVAR is more challenging
and we will adopt the method suggested by \citet{chan2018}. The approach
consists of four steps; 1) obtain a posterior sample via Gibbs sampling,
2) integrate out the time-varying VAR coefficients analytically, 3)
integrate out the stochastic volatility using importance sampling
to obtain the \emph{integrated likelihood}, 4) integrate out the static
parameters using another importance sampler. Since the algorithm makes
use of two nested importance samplers, it is a special case of importance
sampling squared ($IS^{2}$) \citep{tran2013importance}; see Section
\ref{sec:Time-varying-parameter-BVAR} for more details.

There are many alternative estimators that can be used in our approach,
for example, the extension of Chib's estimator to Metropolis-Hastings
sampling \citep{chib2001marginal}, and estimators based on importance
sampling \citep{geweke1999using} or Sequential Monte Carlo \citep{doucet2001sequential}.

All simulation-based estimators: i) give noisy evaluations of the
marginal likelihood, ii) are time-consuming, and iii) have a precision
that is controlled by the user in terms of the number of MCMC or importance
sampling draws. The next section explains how traditional Bayesian
optimization is well suited for points i) and ii), but lacks a mechanism
for exploiting point iii). Taking the user controlled precision into
account brings a new perspective to the problem and we propose a class
of algorithms that handle all three points above.

\section{Bayesian optimization of hyperparameters\label{sec:Bayesian-optimization-for}}

\subsection{Gaussian processes}

Since Bayesian optimization is a relatively unknown method in econometrics,
we give an introduction here to Gaussian processes and their use in
Bayesian optimization.

\begin{figure}[h]
\begin{minipage}[t]{0.4\columnwidth}%
\includegraphics[scale=0.45]{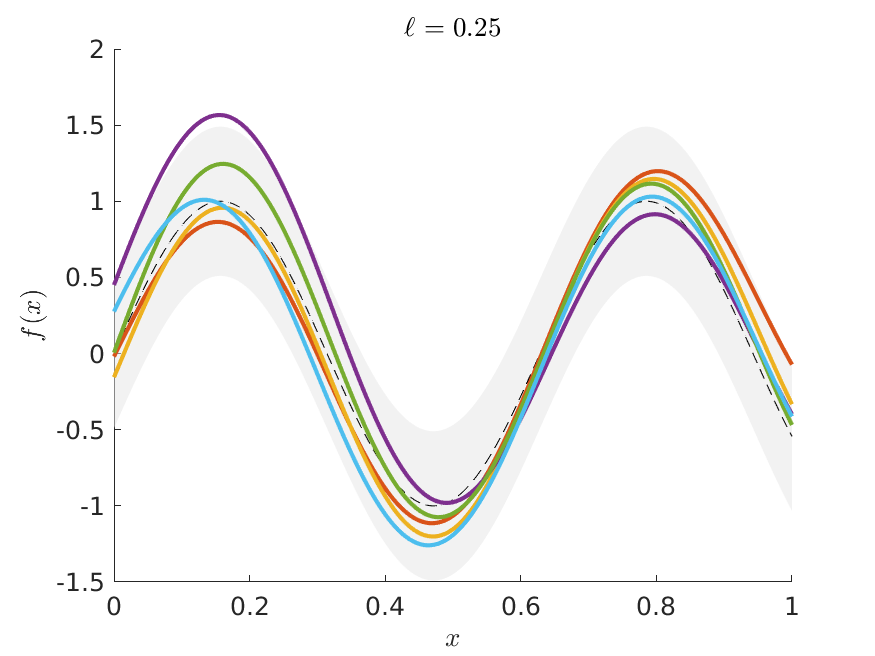}%
\end{minipage}\qquad{}\qquad{}%
\begin{minipage}[t]{0.45\columnwidth}%
\includegraphics[scale=0.45]{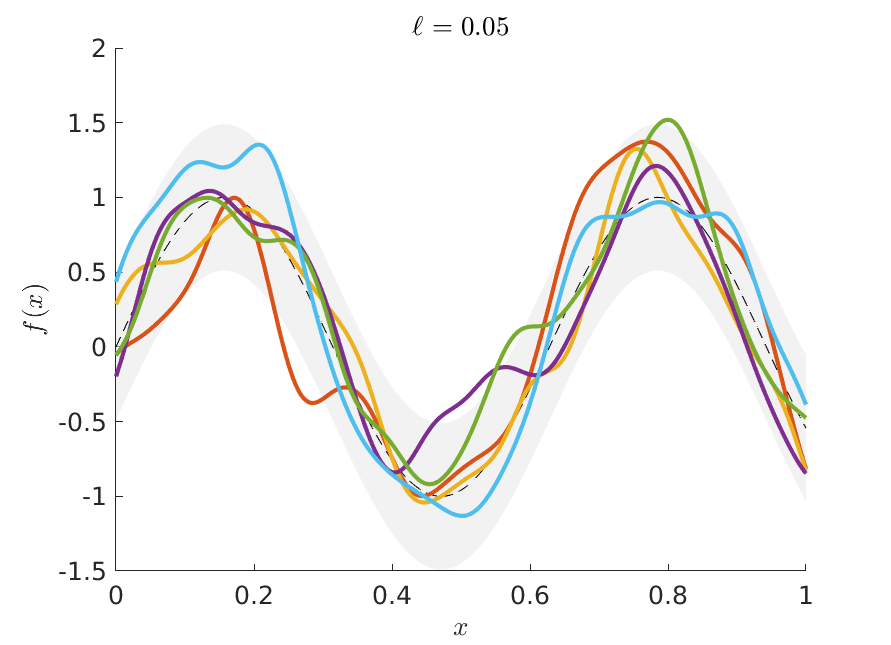}%
\end{minipage}

\caption{Illustration of two Gaussian processes with squared exponential kernel
with different length scales and the same variance $\sigma_{f}^{2}=0.25^{2}$.
The figure shows the prior mean (dashed line) and 95\% probability
intervals (shaded) and five realizations from each process. A smaller
length scale gives more wiggly realizations.\label{fig:GPprior}}
\end{figure}

A Gaussian process (GP) is a (possibly infinite) collection of random
variables such that any subset is jointly distributed according to
a multivariate normal distribution, see e.g. \citet{williams2006gaussian}.
This process, denoted by $f(\mathbf{x})\sim\mathcal{GP}(\mu(\mathbf{x}),k(\mathbf{x},\mathbf{x}'))$,
can be seen as a probability distribution over \emph{functions} $f:\mathcal{X}\to\mathbb{R}$
that is completely specified by its mean function, $\mu(\mathbf{x})\equiv\mathbb{E}f(\mathbf{x})$,
and its covariance function, $\mathbb{C}(f(\mathbf{x}),f(\mathbf{x}'))\equiv k(\mathbf{x},\mathbf{x}')$,
where $\mathbf{x}$ and $\mathbf{x}'$ are two arbitrary input values
to $f(\cdot)$. Note that the covariance function specifies the covariance
between any two \emph{function values}, $f(\mathbf{x}_{1})$ and $f(\mathbf{x}_{2})$.
A popular covariance function is the squared exponential (SE):
\begin{equation}
k(\mathbf{x},\mathbf{x}')=\sigma_{f}\exp\left(-\frac{|\mathbf{x}-\mathbf{x}'|^{2}}{2\ell^{2}}\right),
\end{equation}
where $|\mathbf{x}-\mathbf{x}'|$ is the Euclidean distance between
the two inputs; the covariance function is specified by its two kernel
hyperparameters, the scale parameter $\sigma_{f}>0$ and the length
scale $\ell>0$. The scale parameter $\sigma_{f}$ governs the variability
of the function and the length scale determines how fast the correlation
between two function values taper off with the distance $|\mathbf{x}-\mathbf{x}'|$,
see Figure \ref{fig:GPprior}. The fact that any finite sampling of
function values $\lbrace f(\mathbf{x}_{n})\text{ for }\mathbf{x}_{n}\in\mathcal{X}\rbrace_{n=1}^{N}$
constitutes a multivariate normal distribution on $\mathbb{R}^{N}$
allows for the convenient conditioning and marginalization properties
of the multivariate normal distribution. In particular, this makes
it easy to compute the posterior distribution for the function $f$
at any input $\mathbf{x}_{\star}$.

An increasingly popular alternative to the squared exponential kernel
is the Matérn kernel, see e.g. \citet{matern1960} and \citet{williams2006gaussian}.
The Matérn kernel has an additional hyperparameter, $\nu$>0, in addition
to the length scale $\ell$ and scale $\sigma_{f}$, such that the
process is $k$ times mean square differentiable if and only if $\nu>k$.
Hence, $\nu$ controls the smoothness of the process and it can be
shown that the Matérn kernel approaches the SE kernel as $\nu\rightarrow\infty$
(\citet{williams2006gaussian}). Our approach is directly applicable
for any valid kernel function, but we will use the popular Matérn
$\nu=5/2$ kernel in our applications:

\begin{equation}
k_{\nu=5/2}(r)=\sigma_{f}\left(1+\frac{\sqrt{5}r}{\ell}+\frac{5r^{2}}{3\ell^{2}}\right)\exp\left(-\frac{\sqrt{5}r}{\ell}\right),
\end{equation}
where $r=|\mathbf{x}-\mathbf{x}'|$. The Matérn 5/2 has two continuous
derivatives which is often a requirement for Newton-type optimizers
\citep{snoek2012practical}. The hyperparameters, $\sigma_{f}$ and
$\ell$, are found by maximizing the marginal likelihood, see e.g.
\citet{williams2006gaussian}.

Consider the nonlinear/nonparametric regression model with additive
Gaussian errors
\begin{equation}
y_{i}=f(\boldsymbol{x}_{i})+\epsilon_{i},\quad\epsilon_{i}\overset{\mathrm{iid}}{\sim}N(0,\sigma^{2}),\text{\quad for }i=1,\ldots,n,
\end{equation}
and the prior $f(\mathbf{x})\sim\mathcal{GP}(0,k(\mathbf{x},\mathbf{x}'))$.
Given a dataset with $n$ observations, the posterior distribution
of $f(\boldsymbol{x}_{\star})$ at a new input $\boldsymbol{x}_{\star}$
is \citep{williams2006gaussian}
\begin{align}
f(\boldsymbol{x}_{\star})\,\vert\,y_{1},\ldots,y_{n},\boldsymbol{x}_{1},\ldots,\boldsymbol{x}_{n} & \sim N\left(m(\boldsymbol{x}_{\star}),s^{2}(\boldsymbol{x}_{\star})\right)\nonumber \\
m(\boldsymbol{x}_{\star}) & =k(\boldsymbol{x}_{\star})^{\top}(\boldsymbol{K}(\boldsymbol{X},\boldsymbol{X})+\sigma^{2}I)^{-1}\boldsymbol{y}\nonumber \\
s^{2}(\boldsymbol{x}_{\star}) & =k(\boldsymbol{x}_{\star},\boldsymbol{x}_{\star})-k(\boldsymbol{x}_{\star})^{\top}(\boldsymbol{K}(\boldsymbol{X},\boldsymbol{X})+\sigma^{2}I)^{-1}k(\boldsymbol{x}_{\star}),\label{eq:posteriorGP}
\end{align}
where $\boldsymbol{y}=(y_{1},\ldots,y_{n})^{\top}$, $k(\boldsymbol{x}_{\star})$
is the $n$-vector with covariances between $f$ at the test point
$\boldsymbol{x}_{\star}$ and all other training inputs, $\boldsymbol{K}(\boldsymbol{X},\boldsymbol{X})$
is the $n\times n$ matrix with covariances among the function values
at all $n$ training inputs in $\boldsymbol{X}$. When the errors
are heteroscedastic with variance $\sigma_{i}^{2}$ for the $i$th
observation, the same formulas apply with $\sigma^{2}I$ replaced
by $\mathrm{diag}(\sigma_{1}^{2},\ldots,\sigma_{n}^{2})$.

\subsection{Bayesian optimization}

Bayesian optimization (BO) is an iterative optimization method that
selects new evaluation points using the posterior distribution of
$f$ conditional on the previous function evaluations. More specifically,
BO uses an \emph{acquisition function}, $a(\boldsymbol{x})$, to select
the next evaluation point (\citealp{brochu2010} and \citealp{snoek2012practical}).

An intuitively sensible acquisition rule is to select a new evaluation
point that maximizes the probability of obtaining a higher function
value than the current maximum, i.e. the \emph{Probability of Improvement}
(\emph{PI}):
\begin{equation}
\mathrm{PI}(\mathbf{x})\equiv\mathrm{Pr}(f(\mathbf{x})>f_{max})=\Phi\left(\frac{m(\mathbf{x})-f_{max}}{s(\mathbf{x})}\right),
\end{equation}
where $f_{max}$ is the maximum value of the function obtained so
far. The functions $m(\mathbf{x})$ and $s(\mathbf{x})$ are the posterior
mean and standard deviation of the estimated Gaussian process for
$f$ in the point $\mathbf{x}$, conditional on the available function
evaluations (see (\ref{eq:posteriorGP})), and $\Phi$ denotes the
cumulative standard normal distribution.

The \emph{Expected Improvement} (\emph{EI}) takes also the \emph{size}
of the improvement into consideration:
\begin{equation}
\begin{aligned}\text{EI}(\mathbf{x})= & (m(\mathbf{x})-f_{max})\Phi\left(\frac{m(\mathbf{x})-f_{max}}{s(\mathbf{x})}\right)+s(\mathbf{x})\phi\left(\frac{m(\mathbf{x})-f_{max}}{s(\mathbf{x})}\right),\end{aligned}
\label{EI}
\end{equation}
where $\phi$ denotes the density function of the standard normal
distribution. The first part of (\ref{EI}) is associated with the
size of our predicted improvement and the second part is related to
the uncertainty of our function in that area. Thus, EI incorporates
the trade-off between high expected improvement (exploitation) and
learning more about the underlying function (exploration). Optimization
of the acquisition function, to find the next evaluation point, is
typically a fairly easy task since it is noise-free and cheap to evaluate,
however, it can have multiple optima so some care has to be taken.
An easily implemented solution is to use a regular Newton-type algorithm
initiated with different starting values over the hyperparameter surface.
In this paper we use particle swarm optimization, a global optimization
algorithm implemented in the \texttt{Optim.jl} package in Julia.

\begin{figure}[H]
\begin{centering}
\includegraphics[scale=0.45]{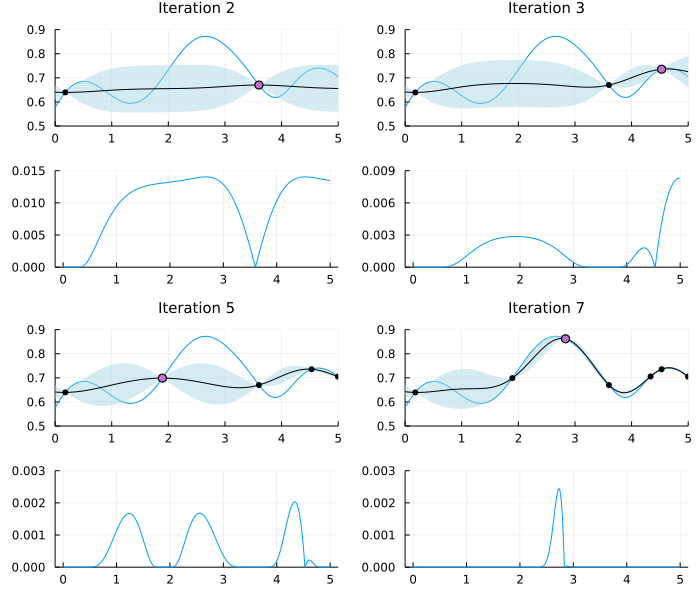}
\par\end{centering}
\caption{Bayesian optimization illustrated with the expected improvements acquisition
strategy. The graphs in Row 1 and 3 depict the posterior distribution
of $f$ and the evaluation points (current evaluation in violet).
Rows 2 and 4 show the corresponding aquisition functions.\label{fig:BO_EI}}
\end{figure}

The exploitation-exploration trade-off is illustrated in Figure \ref{fig:BO_EI},
where the blue line shows the true objective function, the black line
denotes the posterior mean of the GP, and the blue-shaded regions
are 95\% posterior probability bands for $f$. The black (small) dots
are past function evaluations, and the violet (large) dot is the current
evaluation. The first and third row in Figure \ref{fig:BO_EI} show
the true objective function and the learned GP at four different iterations
of the algorithm while the second and fourth row show the EI acquisition
function corresponding to the row immediately above. At Iteration
2 in the top-left corner, we see that the EI acquisition function
(second row) indicates that there is a high expected improvement by
moving to either the immediate left or right of the current evaluation.
At Iteration 5 the EI strategy suggests three regions worth evaluating,
where the two leftmost regions are candidates because of their high
uncertainty. After seven iterations, the algorithm is close to the
global maximum and will now continue a narrow search for the exact
location of the global maximum.

Acquisition rules like PI or EI do not consider that different evaluation
points can be more or less costly. To introduce the notation of cost
into the acquisition strategy, \citet{snoek2012practical} proposed
\emph{Expected Improvement per second}, $\text{EIS}(\mathbf{x})\equiv\text{EI}(\mathbf{x})/c(\mathbf{x})$,
where $c:\mathcal{X}\to\mathbb{R}^{+}$ is a \emph{duration} \emph{function}
that measures the evaluation time at input $\mathbf{x}$ in seconds.
More generally, we can define $a(\mathbf{x})/c(\mathbf{x})$ as an
effort-aware acquisition function. The duration function is typically
unknown and \citet{snoek2012practical} proposed to estimate it alongside
$f$ using an additional Gaussian process for $\log c(\mathbf{x})$.

\subsection{Bayesian optimization with optimized precision}

EIS assumes that the duration (or the cost) of function evaluations
are unknown, but \emph{fixed} for a given input $\mathbf{x}$; once
we visit $\mathbf{x}$, the cost of the function estimate $\hat{f}(\mathbf{x})$
is given. However, the user can often \emph{choose} the duration spent
to obtain a certain precision in the estimate; for example by increasing
the number of MCMC iterations when the marginal likelihood is estimated
by MCMC. This novel perspective opens up for strategies that not only
optimize for the next evaluation point, but also optimize over the
computational resources, or equivalently, the precision of the estimate
$\hat{f}(\mathbf{x})$. We formally extend BO by modeling the function
evaluations with a heteroscedastic GP
\begin{align}
\hat{f}(\mathbf{x}) & =f(\mathbf{x})+\epsilon,\:\epsilon\sim N(0,\sigma^{2}(\mathbf{x},G))\label{eq:GPforNoisyEvals}\\
f & \sim\mathcal{GP}(\mu(\mathbf{x}),k(\mathbf{x},\mathbf{x}')),\nonumber 
\end{align}
where the noise variance $\sigma^{2}(\mathbf{x},G)$ is now an explicit
function of the number of MCMC iterations, $G$, or some other duration
measure. Hence the user can now choose both \emph{where} to place
the next evaluation and the \emph{effort} spent in computing it by\emph{
}maximizing
\begin{equation}
\tilde{a}(\mathbf{x},G)\equiv a(\mathbf{x})/G,
\end{equation}
with respect to both $\mathbf{x}$ and $G$, where $a(\mathbf{x})$
is a baseline acquisition function, for example EI.

A complication with maximization of $\tilde{a}(\mathbf{x},G)$ is
that while we typically know that $\sigma(\mathbf{x},G)=O(1/\sqrt{G})$
in Monte Carlo or MCMC, the exact numerical standard error depends
on the integrated autocorrelation time (IACT) of the MCMC chain. Note
that the evaluation points can, for example, be hyperparameters in
the prior, where different values can give rise to varying degrees
of well-behaved posteriors, so we can not expect the IACT to be constant
over the hyperparameter space, hence the explicit dependence on $\boldsymbol{x}$
in $\sigma^{2}(\mathbf{x},G)$. Rather than maximizing $\tilde{a}(\mathbf{x},G)$
with respect to both $\mathbf{x}$ and $G$ directly, we propose to
implement the algorithm in an alternative way that achieves a similar
effect. The approach includes stopping the evaluation early whenever
the function evaluation turns out to be hopelessly low with a low
probability of improvement over the current $f_{max}$.

For a given $\mathbf{x}$ we let $G$ increase, in batches of a fixed
size, until
\begin{equation}
\text{PI}(\mathbf{x})\equiv\Phi\left(\frac{\hat{m}^{(g)}(\mathbf{x})-f_{max}}{s^{(g)}(\mathbf{x})}\right)<\alpha,
\end{equation}
for some small value, $\alpha$, or until $G$ reaches a predetermined
upper bound, $\bar{G}$; here $\hat{m}^{(g)}(\mathbf{x})\text{ and }s^{(g)}(\mathbf{x})$
denotes the posterior mean and standard deviation of the GP evaluated
at $\mathbf{x}$ after $g$ MCMC iterations. Note that both the posterior
mean $m(\mathbf{x})$ and standard deviation $s(\mathbf{x})$ are
functions of the noise variance, which in turn is a function of $G$.
The posterior distribution for $f(\mathbf{x})$ is hence continuously
updated as $G$ grows until $1-\alpha$ of the posterior mass in the
GP for $f(\mathbf{x})$ is concentrated below $f_{max}$, at which
point the evaluation stops. The optimization is insensitive to the
choice of $\alpha$, as long as it is a relatively small number. We
now propose to maximize the following acquisition function based on
early stopping
\begin{equation}
\tilde{a}_{\alpha}(\mathbf{x})=a(\mathbf{x})/\hat{G}_{\alpha}(\mathbf{x}),\label{eq:earlyStoppingRule}
\end{equation}
where $\hat{G}_{\alpha}(\mathbf{x})$ is a \emph{prediction} of the
number of MCMC draws needed at $\mathbf{x}$ before the evaluation
stops, with the probability $\alpha$ as the threshold for stopping.
We emphasize that early stopping is here used in a subtle way, not
only as a simple rule to short-circuit useless computations, but also
in the \emph{planning} of future computations; the mere possibility
of early stopping can make the algorithm try an $\mathbf{x}$ which
does not have the highest $a(\mathbf{x})$, but which is expected
to be cheap and is therefore worth a try. This effect that comes via
$\sigma^{2}(G)$ is not present in the EIS of \citet{snoek2012practical}
where the cost is fixed and is not influenced by the probability model
on $f$.

Although one can use any model to predict $G$, we will here fit a
GP regression model to the logarithm of the number of MCMC draws,
$\log G_{j}$ for $j=1,...,J$ in the $J$ previous evaluations
\begin{align}
\log G_{j} & =h(\mathbf{z}_{j})+\varepsilon_{j},\:\varepsilon_{j}\overset{iid}{\sim}N(0,\psi^{2})\nonumber \\
h & \sim\mathcal{GP}(m_{G}(\mathbf{z}),k_{G}(\mathbf{z},\mathbf{z}')),\label{eq:effortPrediction}
\end{align}
where $\mathbf{z}_{j}$ is a vector of covariates. The hyperparameters,
$\mathbf{x}_{1:J}$, themselves may be used as predictors of $\hat{G}(\mathbf{x})$,
but also $D(\mathbf{x})=\hat{m}(\mathbf{x})-f_{max}$ and $s(\mathbf{x})$
are likely to have predictive power for $G$, as well as $u(\mathbf{x})=(\hat{m}(\mathbf{x})-f_{max})/s(\mathbf{x})$.
We will use $\mathbf{z}_{j}=\left(\mathbf{x}_{j},D^{(j)}(\mathbf{x}_{j}),s^{(j)}(\mathbf{x}_{j}),u^{(j)}(\mathbf{x}_{j})\right)$
in our applications, where the superscript over $j$ denotes the BO
iteration. The prediction for $G$ is taken to be $\hat{G}=\text{exp}\left(m_{G}\left(\mathbf{z}\right)\right)$,
which corresponds to the median of the log-normal posterior for $G$.

We will use the term \emph{Bayesian Optimization with Optimized Precision}
(BOOP) for BO methods that optimize $\tilde{a}_{\alpha}(\mathbf{x})$
in (\ref{eq:earlyStoppingRule}), and more specifically BOOP-EI when
EI is used as the baseline acquisition function, $a(\mathbf{x})$.
The whole procedure is described in Algorithm \ref{FullAlg}. \begin{algorithm}[H] \caption{Bayesian Optimization with Optimized Precision (BOOP)}\label{FullAlg} 

\vspace{0.2cm}
\textbf{input}
\begin{itemize}
\item estimator $\hat f(\mathbf{x})$ of the $f(\mathbf{x})$ to be maximized, and its standard error function $\sigma(G)$.
\item $j_0$ initial points $\mathbf{x}_{1:j_0}\equiv(\mathbf{x}_1,\ldots,\mathbf{x}_{j_0})$, a vector of corresponding function estimates, $\hat{f}(\mathbf{x}_{1:j_0})$, and standard errors $\sigma^{2}(G_{1:j_0})$.
\item baseline acquisition function $a(\mathbf{x})$, and early stopping thresholding probability $\alpha$.
\end{itemize}


\textbf{for $j$ from $j_0+1$ until convergence do:} 
\begin{enumerate} 
\item[a)] Fit the heteroscedastic GP for $f$ based on past evaluations
\begin{align*} 
\hat{f}(\mathbf{x}_{1:(j-1)})&=f(\mathbf{x}_{1:(j-1)})+\epsilon, \hspace{0.3cm}\epsilon\sim N(0,\Sigma_{1:(j-1)}) \\
f(\mathbf{x})&\sim \mathcal{GP}(m(\mathbf{x}),k(\mathbf{x},\mathbf{x}')), 
\end{align*}
where $\Sigma_{1:(j-1)} \equiv \mathrm{Diag}(\sigma^{2}(G_1),\ldots,\sigma^{2}(G_{j-1}))$.
\item[b)] Fit the GP for $\log G$ based on past evaluations
\begin{align*} 
\log G_{1:(j-1)} &=h(\mathbf{z}_{1:(j-1)})+\varepsilon, \hspace{0.3cm} \varepsilon {\sim}N(0,\psi^{2}\mathbf{I}) \\ 
h(\mathbf{z}) &\sim \mathcal{GP}(m_G(\mathbf{z}),k_G(\mathbf{z},\mathbf{z}')),
\end{align*}
where the elements of $\mathbf{z}$ are functions of $\mathbf{x}$. Return the point prediction $\hat G_\alpha(\mathbf{x})$.
\item[c)] Maximize $\tilde a_{\alpha}(\mathbf{x})=a(\mathbf{x})/\hat G_\alpha(\mathbf{x})$ to select the next point, $\mathbf{x}_j$.
\item[d)] Compute $\hat f (\mathbf{x}_j)$ and $\sigma^{2}(G_j)$ by early stopping at thresholding probability $\alpha$.
\item[e)] Update the datasets in a) with $(\mathbf{x}_j,\hat f (\mathbf{x}_j),\sigma^2(G_i))$ and in b) with $(\mathbf{z}_j,\log G_j)$.
\end{enumerate} 
\end{algorithm} 
 Note that (\ref{eq:GPforNoisyEvals}) assumes that $\hat{f}(\mathbf{x})$
is an unbiased estimator at any $\mathbf{x}$. This can be ensured
by using enough MCMC/importance sampling draws in the first marginal
likelihood evaluation of BOOP. We performed a small simulation exercise
that shows that the Chib estimator is approximately unbiased after
a small number of iterations for the medium-sized VAR model in \ref{subsec:Results-for-medium-scale}.
As expected, we had to use more initial draws in the large-scale VAR
in Section \ref{subsec:Results-for-the-largescale} and the time-varying
parameter VAR with stochastic volatility in Section \ref{sec:Time-varying-parameter-BVAR}
to drive down the bias. See also Section \ref{sec:Conclusion} for
some ideas on how to extend BOOP to estimators where the bias is still
sizeable in large MCMC samples. 

\begin{figure}[H]
\begin{centering}
\includegraphics[scale=0.45]{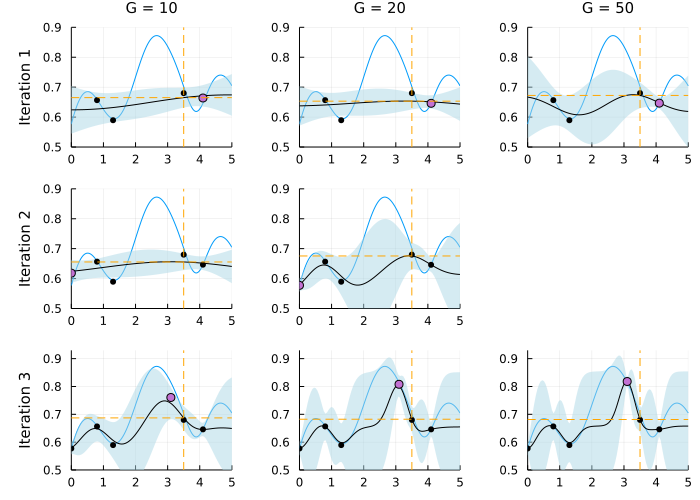}
\par\end{centering}
\caption{\label{fig:BOearlyStop}Illustration of BOOP-EI implemented with early
stopping. The row corresponds to iterations in the algorithm and the
columns to different MCMC sample sizes. The blue line is the true
$f$, the shaded regions are 95\% posterior probability bands for
$f$ based on the (noisy) evaluations (black for past and violet for
current). The orange crosshair marks the current maximum; see the
text for details.}
\end{figure}

Figure \ref{fig:BOearlyStop} illustrates the early stopping part
of BOOP in a toy example. The first row illustrates the first BOOP
iteration and the columns show increasingly larger MCMC sample sizes
($G$). We can see that the 95\% posterior interval after $G=10$
MCMC draws at the current $\mathbf{x}$ includes $f_{max}^{(1)}$
(dotted orange line), the highest posterior mean of the function values
observed so far; it therefore worthwhile to increase the number of
simulations for this $\mathbf{x}$. Moving one graph to the right
we see that after $G=20$ simulations the 95\% posterior interval
still includes $f_{max}^{(1)}$, and we move one more graph to the
right for $G=50$. Here we conclude that the sampled point is almost
certainly not an improvement and we move on to a new evaluation point.
The new evaluation point is found by maximizing the BOOP-EI acquisition
function in (\ref{eq:earlyStoppingRule}) with updated effort prediction
function $\hat{G}(\boldsymbol{z})$ in Equation \ref{eq:effortPrediction}
and is depicted by the violet dot in leftmost graph in the second
row of Figure \ref{fig:BOearlyStop}. Following the progress in the
second row, we see that it takes only $G=20$ samples to conclude
that the function value is almost certainly lower than the current
maximum of the posterior mean at the second BO iteration. Finally,
in the third row, we can see that the point is sampled with high variance
at the beginning, but as we increase $G$ it becomes clear that this
$\mathbf{x}$ is indeed an improvement.

The code used in this paper is written in the Julia computing language
\citep{bezanson2017julia}, making use of the \texttt{GaussianProcesses.jl}\emph{
}package for estimation of the Gaussian processes and the \texttt{Optim.jl}
package for the optimization of the acquisition functions.

\section{Simulation experiment\label{sec:Simulation-experiment}}

\subsection{Simulation setup}

We will here assess the performance of the proposed BOOP-EI algorithm
in a simulation experiment for the optimization of a non-linear function
in a single variable. The simple one-dimensional setup is chosen for
presentational purposes, and more challenging higher-dimensional settings
are likely to show even larger advantages of our method compared to
regular Bayesian optimization.

\begin{figure}[H]
\begin{centering}
\includegraphics[scale=0.4]{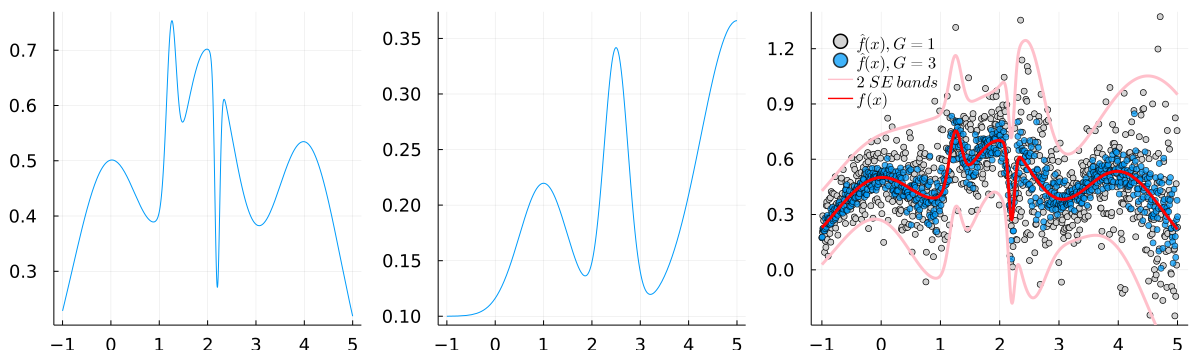}
\par\end{centering}
\caption{The function $f$ that we want to maximize (left) and the function
$g$ that controls the sampling variance over $x$ (middle). The figure
to the right shows estimates for $G=1$ samples (grey dots), $G=3$
samples (blue dots), the mean function (red line) and $2$ standard
deviation error bands (pink lines). \label{fig:Simul_func}}
\end{figure}

The function to be maximized is

\begin{equation}
\begin{aligned}f(x) & =N(x|0,0.8^{2})+N(x|4,0.75^{2})+N(x|0,0.8^{2})+N(x|2,0.6^{2})+\\
 & +0.05N(x|2.2,0.05^{2})+0.075N(x|1.25,0.1^{2}),
\end{aligned}
\label{eq:sim_model}
\end{equation}
where $N(x|\mu,\sigma^{2})$ denotes the density function of a $N(\mu,\sigma^{2})$
variable; the function is plotted in Figure \ref{fig:Simul_func}
(left). We further assume that $f(x)$ can be estimated at any $x$
from $G$ noisy evaluations by a simple Monte Carlo average
\begin{equation}
\hat{f}^{(G)}(x)=f(x)+\epsilon,\:\epsilon\sim N\left(0,\frac{g^{2}(x)}{G}\right)
\end{equation}
where $g(x)=0.1+0.15N(x|1,0.5^{2})+0.15N(x|2.5,0.25^{2})+0.5N(x|5,0.75^{2})$
is a heteroscedastic standard deviation function that mimics the real
world case where the variability of the marginal likelihood estimate
varies over the space of hyperparameters; $g(x)$ is plotted in Figure
\ref{fig:Simul_func} (middle). Figure \ref{fig:Simul_func} (right)
illustrates the noisy function evaluations (gray points) and the effect
of Monte Carlo averaging $G=3$ evaluations for a given $x$ (blue
points). We assume for simplicity here that once the algorithm decides
to visit an $x$ it will get access to a noise-free evaluation of
the standard error of the estimator.

\subsection{Illustration of a single run of the algorithms}

\begin{figure}[H]
\begin{centering}
\begin{minipage}[t]{0.48\columnwidth}%
\begin{center}
\includegraphics[scale=0.38]{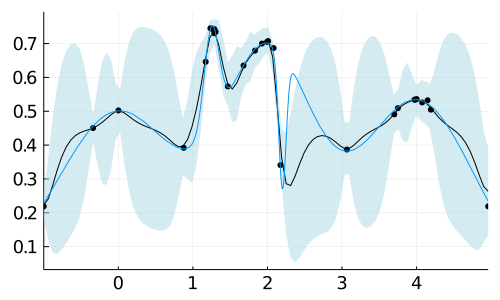}
\par\end{center}%
\end{minipage}\quad{}%
\begin{minipage}[t]{0.48\columnwidth}%
\begin{center}
\includegraphics[scale=0.38]{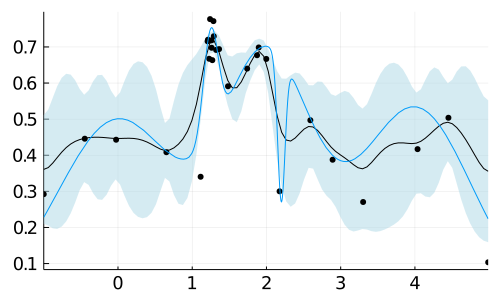}
\par\end{center}%
\end{minipage}
\par\end{centering}
\caption{GP posterior after 25 Bayesian optimization iterations using BO-EI
(left) and BOOP-EI (right) to optimize $f$. The black lines are the
posterior means of $f(x).$\label{fig:BOOPvsEI}}
\end{figure}

Before we conduct a small Monte Carlo experiment it is illustrative
to look at the results from a single run of the EI and BOOP-EI algorithms.
Figure \ref{fig:BOOPvsEI} highlights the difference between the algorithms
by showing the GP posterior after 25 Bayesian optimization runs. The
EI algorithm has clearly wasted computations to needlessly drive down
the uncertainty at (useless) low function values, while BOOP-EI tolerates
larger uncertainty in such function locations.

\subsection{Simulation study}

We now compare the performance of a standard EI approach using a heteroscedastic
GP with our BOOP-EI approach in a small simulation study. The methods
will be judged by their ability to find the optimum using as few evaluations
as possible. The performance will be evaluated by a Monte Carlo study
where we simulate $1000$ replications from each model under each
simulation scenario. We investigate Bayesian optimization with EI
using $100$ and $500$ samples in each iteration, BOOP-EI is allowed
to stop the sampling at any time before the number of samples for
the EI is reached. In each simulation we set an upper bound of $120$
Bayesian optimization iterations.

\begin{figure}[H]
\begin{centering}
\textasciiacute{}
\par\end{centering}
\begin{centering}
\begin{minipage}[t]{0.5\columnwidth}%
\begin{center}
\includegraphics[scale=0.34]{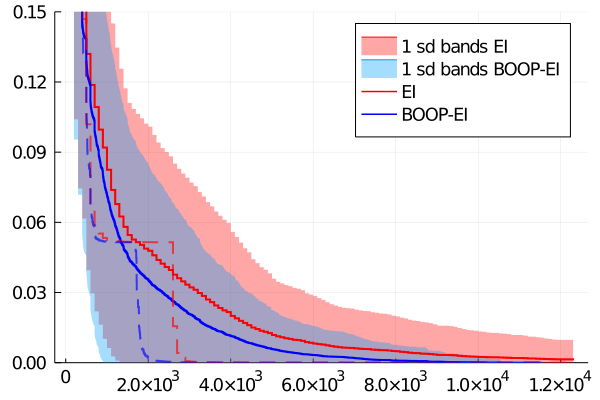}
\par\end{center}%
\end{minipage}%
\begin{minipage}[t]{0.5\columnwidth}%
\begin{center}
\includegraphics[scale=0.34]{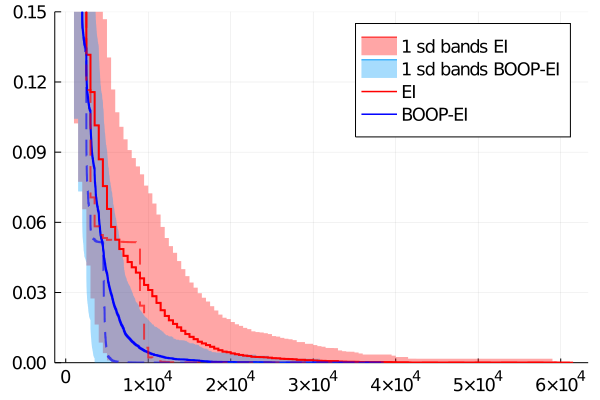}
\par\end{center}%
\end{minipage}
\par\end{centering}
\caption{The evolution of $f_{max}-\max f(x)$, i.e. the difference between
the current maximum $f_{max}$ and the true maximum of the function
(vertical axis), as a function of the \emph{total} number of the MCMC
draws consumed up to the current BO/BOOP iteration. Note that the
number of MCMC draws per BOOP iteration is variable due to early stopping.
The lefthand graph uses a maximum of $100$ MCMC draws per BO iteration
while the righthand graph uses $500$ draws. The shaded areas are
the one standard deviation probability bands in the distribution of
$f_{max}-\max f(x)$ over the replicate runs, and the solid and dashed
lines are the mean and median, respectively. \label{evolution_max}}
\end{figure}

We can see from Figure \ref{evolution_max} that BOOP finds the maximum
by using fewer total number of samples in both scenarios and that
the difference increases with the number of samples used when forming
the estimator. We can also see from the median $f_{max}$ that both
algorithms can get stuck for a while at the second greatest local
maximum (which is approximately $.05$ lower than the global maximum).
However, BOOP gets out of the local optimum faster since it has the
option to try cheap noisy evaluations and will therefore explore other
parts of the function earlier than basic BO-EI. This effect seems
to increase as we allow for a higher number of samples since it lowers
the relative price of cheaper evaluations.

\section{Application to the steady-state BVAR\label{sec:Empirical-applications}}

In this section, we use BOOP-EI to estimate the prior hyperparameters
of the steady-state BVAR of \citet{villani2009}. \citet{giannone2015}
show that finding the right values for the hyperparameters in BVARs
can significantly improve forecasting performance. Moreover, \citet{banbura}
show that different degree of shrinkage (controlled by the hyperparameters)
is necessary under different model specifications.

\subsection{The steady-state BVAR}

The steady-state BVAR model of \citet{villani2009} is given by:

\begin{equation}
\begin{aligned}\Pi(L)(\mathbf{y}_{t}-\Psi\mathbf{x}_{t}) & =\mathbf{\varepsilon}_{t},\qquad\text{}\mathbf{\varepsilon}_{t}\overset{\text{iid}}{\sim}N(\mathbf{0},\Sigma),\end{aligned}
\end{equation}
where $E[\mathbf{y}_{t}]=\Psi\mathbf{x}_{t}$. In particular, if we
assume that $\mathbf{x}_{t}=1$ for all $t,$ then $\Psi$ has the
interpretation as the overall mean of the process. We take the prior
distribution to be:
\begin{equation}
\begin{aligned}p(\Sigma)\sim & \Sigma^{-(n+1)/2}\\
vec(\Pi)\sim & N(\underline{\boldsymbol{\theta}}_{\Pi},\underline{\Omega}_{\Pi})\\
\Psi\sim & N(\underline{\boldsymbol{\theta}}_{\Psi},\underline{\Omega}_{\Psi}),
\end{aligned}
\end{equation}
where $\underline{\boldsymbol{\theta}}_{\Psi}\text{ and }\underline{\Omega}_{\Psi}$
are the mean and covariance matrix for the steady states. The prior
covariance matrix for the dynamics, $\underline{\Omega}_{\Pi}$, is
constructed using 
\begin{equation}
\underline{\omega}_{ii}=\begin{cases}
\frac{\theta_{1}^{2}}{(l^{\theta_{3}})^{2}},\text{ for own lag }l\text{ of variable }r,\ i=(l-1)n+r,\\
\frac{(\theta_{1}\theta_{2}s_{r})^{2}}{(l^{\theta_{3}}s_{j})^{2}},\text{ for cross-lag }l\text{ of variable }r\neq j,\ i=(l-1)n+j,
\end{cases}
\end{equation}
 where $\underline{\omega}_{ii}$ is the diagonal elements of $\underline{\Omega}_{\Pi}$.
We also assume prior independence, following \citet{villani2009}.
The hyperparameters that we optimize over are: the \emph{overall-shrinkage
parameter }$\theta_{1}$, the \emph{cross-lag shrinkage} $\theta_{2}$,\emph{
}and the \emph{lag-decay parameter} $\theta_{3}$\emph{.}

The posterior distribution of the steady-state BVAR model parameters
can be sampled with a simple Gibbs sampling scheme \citep{villani2009}.
The marginal likelihood, together with its empirical standard error,
can be estimated by the method in \citet{chib1995marginal}.

\subsection{Data, prior and model settings}

Table \ref{tab:data} describes the data used in our applications
which are also used in \citet{giannone2015}. It contains 23 macroeconomic
variables for which two subsets are selected to represent a medium-sized
model with 7 variables and a large model that contains 22 of the variables
(real investment is excluded). Before the analysis, the consumer price
index and the five-year bond rate were transformed from monthly to
quarterly frequency. All series are transformed such that they become
stationary according to the augmented Dickey-Fuller test. This is
necessary for the data to be consistent with the prior assumption
of a steady-state. The number of lags is chosen according to the HQ-criteria,
\citet{hannan1979determination} and \citet{quinnMulti}. This resulted
in $p=2$ lags for the medium-sized model which we also use for the
large model.

We set the prior mean of the coefficient matrix, $\Pi$, to values
that reflect some persistence on the first lag, but also that all
the time series are stationary; e.g. the prior mean on the first lag
of the FED interest rate and the GDP-deflator is set to 0.6, while
others are set to zero in the medium-sized model. Lags longer than
1 and cross-lags all have zero prior means. The priors for the steady-states
are set informative to the values listed in Table \ref{tab:data},
these values follow suggestions from the literature for most variables,
see e.g. \citet{louzis2019} and \citet{osterholm2012limited}. There
were a few variables where we could not find theoretical values for
either the mean or the standard deviation, in those cases, we set
them close to their empirical counterparts.

\begin{table}[H]
\begin{centering}
\begin{tabular}{lccccc}
\multicolumn{6}{c}{{\footnotesize{}Variable names and transformations}}\tabularnewline
\hline 
\multirow{2}{*}{{\footnotesize{}Variables}} & {\footnotesize{}Mnemonic} & \multirow{2}{*}{{\footnotesize{}Transform}} & \multirow{2}{*}{{\footnotesize{}Medium}} & \multirow{2}{*}{{\footnotesize{}Freq.}} & \multirow{2}{*}{{\footnotesize{}Prior}}\tabularnewline
 & {\footnotesize{}(FRED)} &  &  &  & \tabularnewline
\hline 
\hline 
{\footnotesize{}Real GDP} & {\footnotesize{}GDPC1} & {\footnotesize{}$400\times\text{diff-log}$} & {\footnotesize{}x} & {\footnotesize{}Q} & {\footnotesize{}(2.5;3.5)}\tabularnewline
{\footnotesize{}GDP deflator} & {\footnotesize{}GDPCTPI} & {\footnotesize{}$400\times\text{diff-log}$} & {\footnotesize{}x} & {\footnotesize{}Q} & {\footnotesize{}(1.5;2.5)}\tabularnewline
{\footnotesize{}Fed funds rate} & {\footnotesize{}FEDFUNDS} & {\footnotesize{}-} & {\footnotesize{}x} & {\footnotesize{}Q} & {\footnotesize{}(4.3,5.7)}\tabularnewline
{\footnotesize{}Consumer price index} & {\footnotesize{}CPIAUCSL} & {\footnotesize{}$400\times\text{diff-log}$} &  & {\footnotesize{}M} & {\footnotesize{}(1.5;2.5)}\tabularnewline
{\footnotesize{}Commodity prices} & {\footnotesize{}PPIACO} & {\footnotesize{}$400\times\text{diff-log}$} &  & {\footnotesize{}Q} & {\footnotesize{}(1.5;2.5)}\tabularnewline
{\footnotesize{}Industrial production} & {\footnotesize{}INDPRO} & {\footnotesize{}$400\times\text{diff-log}$} &  & {\footnotesize{}Q} & {\footnotesize{}(2.3;3.7)}\tabularnewline
{\footnotesize{}Employment} & {\footnotesize{}PAYEMS} & {\footnotesize{}$400\times\text{diff-log}$} &  & {\footnotesize{}Q} & {\footnotesize{}(1.5;2.5)}\tabularnewline
{\footnotesize{}Employment, service sector} & {\footnotesize{}SRVPRD} & {\footnotesize{}$400\times\text{diff-log}$} &  & {\footnotesize{}Q} & {\footnotesize{}(2.5;3.5)}\tabularnewline
{\footnotesize{}Real consumption} & {\footnotesize{}PCECC96} & {\footnotesize{}$400\times\text{diff-log}$} & {\footnotesize{}x} & {\footnotesize{}Q} & {\footnotesize{}(2.3;3.7)}\tabularnewline
{\footnotesize{}Real investment} & {\footnotesize{}GPDIC1} & {\footnotesize{}$400\times\text{diff-log}$} & {\footnotesize{}x} & {\footnotesize{}Q} & {\footnotesize{}(1.5;4.5)}\tabularnewline
{\footnotesize{}Real residential investment} & {\footnotesize{}PRFIx} & {\footnotesize{}$400\times\text{diff-log}$} &  & {\footnotesize{}Q} & {\footnotesize{}(1.5;4.5)}\tabularnewline
{\footnotesize{}Nonresidential investment} & {\footnotesize{}PNFIx} & {\footnotesize{}$400\times\text{diff-log}$} &  & {\footnotesize{}Q} & {\footnotesize{}(1.5;4.5)}\tabularnewline
{\footnotesize{}Personal consumption} & \multirow{2}{*}{{\footnotesize{}PCECTPI}} & \multirow{2}{*}{{\footnotesize{}$400\times\text{diff-log}$}} & \multirow{2}{*}{} & \multirow{2}{*}{{\footnotesize{}Q}} & \multirow{2}{*}{{\footnotesize{}(1.5;4.5)}}\tabularnewline
{\footnotesize{}expenditure, price index} &  &  &  &  & \tabularnewline
{\footnotesize{}Gross private domestic} & \multirow{2}{*}{{\footnotesize{}GPDICTPI}} & \multirow{2}{*}{{\footnotesize{}$400\times\text{diff-log}$}} & \multirow{2}{*}{} & \multirow{2}{*}{{\footnotesize{}Q}} & \multirow{2}{*}{{\footnotesize{}(1.5;4.5)}}\tabularnewline
{\footnotesize{}investment, price index} &  &  &  &  & \tabularnewline
{\footnotesize{}Capacity utilization} & {\footnotesize{}TCU} & {\footnotesize{}-} &  & {\footnotesize{}Q} & {\footnotesize{}(79.3;80.7)}\tabularnewline
{\footnotesize{}Consumer expectations} & {\footnotesize{}UMCSENTx} & {\footnotesize{}diff} &  & {\footnotesize{}Q} & {\footnotesize{}(-0.5, 0.5)}\tabularnewline
{\footnotesize{}Hours worked} & {\footnotesize{}HOANBS} & {\footnotesize{}$400\times\text{diff-log}$} & {\footnotesize{}x} & {\footnotesize{}Q} & {\footnotesize{}(2.5;3.5)}\tabularnewline
{\footnotesize{}Real compensation/hour} & {\footnotesize{}AHETPIx} & {\footnotesize{}$400\times\text{diff-log}$} & {\footnotesize{}x} & {\footnotesize{}Q} & {\footnotesize{}(1.5;2.5)}\tabularnewline
{\footnotesize{}One year bond rate} & {\footnotesize{}GS1} & {\footnotesize{}diff} &  & {\footnotesize{}Q} & {\footnotesize{}(-0.5;0.5)}\tabularnewline
{\footnotesize{}Five years bond rate} & {\footnotesize{}GS5} & {\footnotesize{}diff} &  & {\footnotesize{}M} & {\footnotesize{}(-0.5,0.5)}\tabularnewline
{\footnotesize{}SP 500} & {\footnotesize{}S\&P 500} & {\footnotesize{}$400\times\text{diff-log}$} &  & {\footnotesize{}Q} & {\footnotesize{}(-2,2)}\tabularnewline
{\footnotesize{}Effective exchange rate} & {\footnotesize{}TWEXMMTH} & {\footnotesize{}$400\times\text{diff-log}$} &  & {\footnotesize{}Q} & {\footnotesize{}(-1;1)}\tabularnewline
{\footnotesize{}M2} & {\footnotesize{}M2REAL} & {\footnotesize{}$400\times\text{diff-log}$} &  & {\footnotesize{}Q} & {\footnotesize{}(5.5;6.5)}\tabularnewline
\hline 
\end{tabular}\caption{Data Description\label{tab:data}}
\par\end{centering}
\raggedright{}{\footnotesize{}The table shows the 23 US macroeconomic
time series from the FRED database used in the empirical analysis.
The column named Prior contains the steady-state mean $\pm$ one standard
deviation.}{\footnotesize\par}
\end{table}

\subsection{Experimental setup}

We consider three competing optimization strategies: (I) an exhaustive
grid-search, (II) Bayesian optimization with the EI acquisition function
(BO-EI), and (III) our BOOP-EI algorithm. In each approach, we use
the restrictions $\theta_{1}\in(0,5),\theta_{2}\in(0,1)\text{, and }\theta_{3}\in(0,5)$.
In the grid-search, $\theta_{1}\text{ and }\theta_{2}$ move in steps
of $0.05$ and $\theta_{3}$ moves in steps of $0.1$, yielding in
total 100000 marginal likelihood evaluations. For the Bayesian optimization
algorithm, we set the number of evaluations to 250, and we use three
random draws as initial values for the GPs.

\begin{figure}[H]
\begin{centering}
\includegraphics[scale=0.4]{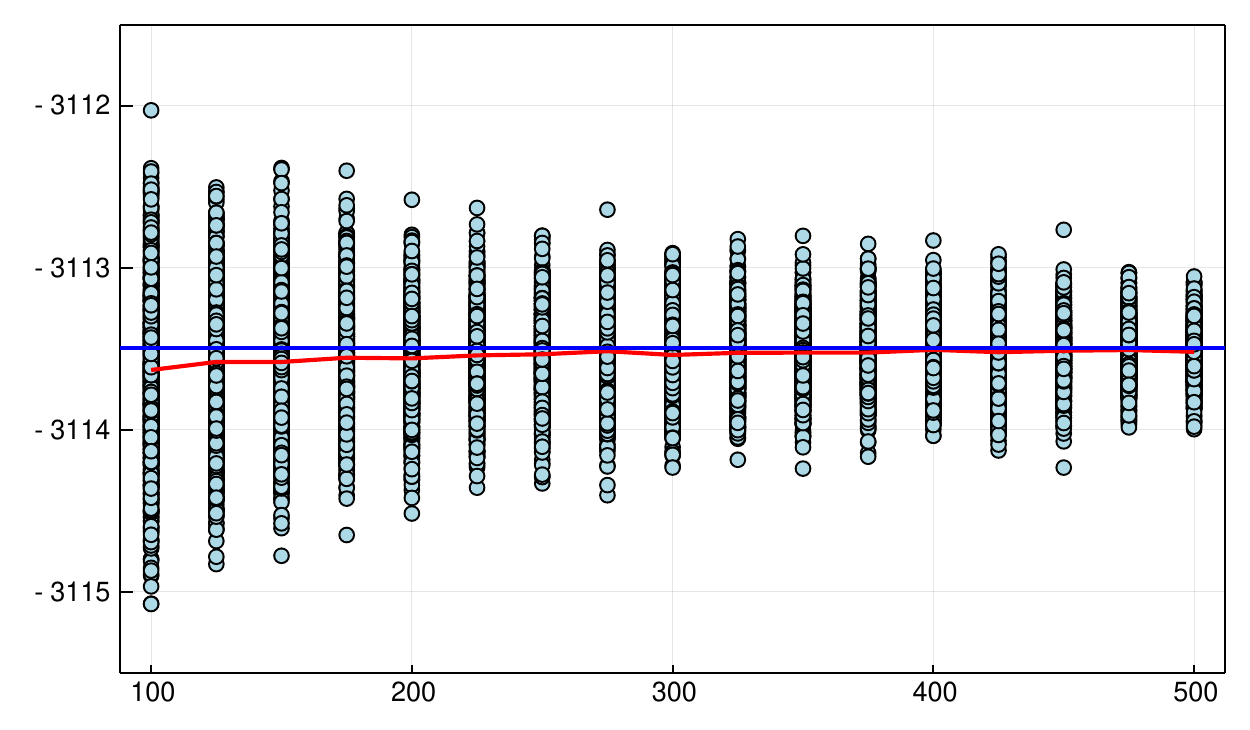}
\par\end{centering}
\caption{Unbiasedness of Chib's log marginal likelihood estimator in the steady-state
BVAR application. The horizontal axis denotes the number of MCMC draws
(excluding 50 observations as burn-in), the blue dots are draws from
the sampling distribution of Chib's estimator for a given MCMC sample
size. The red line represents the mean of the draws and the blue line
represents the true log marginal likelihood, obtained from 100 000
MCMC iterations with 5000 as a burn-in.\label{fig:Convergence-of-Chib's}}
\label{MCMC_Conv-1}
\end{figure}

For strategies (I) and (II) we use a total of 10000 Gibbs iterations
with 1000 as a burn-in sample in each model evaluation. For (III)
we first draw 1100 Gibbs samples where we discard the first 1000 as
burn-in and use the rest to calculate the probability of improvement
PI, to ensure that the estimated marginal likelihood will be approximately
unbiased; Figure \ref{fig:Convergence-of-Chib's} shows that Chib's
estimate is unbiased already after a few hundred samples. If $\text{PI}<\alpha$
we stop early and move on to the next BO iteration, otherwise we generate
a new batch (of size 100) of Gibbs samples and again check the PI
criteria. The total number of Gibbs iterations will therefore vary
between 1100 and 10 000 in each of the 250 BOOP-iterations for the
medium-sized model. Note that Chib's estimator uses an estimate of
the parameter for the so-called reduced Gibbs sampler run. This point
estimate should preferably have a high posterior density for efficiency
reasons, see \citet{chib1995marginal}. The medium-sized model uses
only 100 posterior samples to obtain high-density parameters for calculating
Chib's log marginal likelihood, which is enough in our set-up. For
the large model, we use $5000$ burn-in samples and $500$ simulations
in the first batch, which gives between $5500$ and $10000$ MCMC
iterations per evaluation point. The $5000$ burn-in is likely to
be excessive but is used to be conservative; a small number would
make BOOP even faster in comparison to regular BO. The application
is robust to the choice of $\alpha$, as long as it is a reasonably
low number, in this study we use $\alpha=0.001$.

For comparison, we will also use the standard values of the hyperparameters
used in e.g. the BEAR-toolbox, \citet{bear}, $\theta_{1}=0.1,\theta_{2}=0.5,\text{ and }\theta_{3}=1$,
as a benchmark. The methods are compared with respect to i) the obtained
marginal likelihood, and ii) how much computational resources were
spent in the optimization.

\subsection{Results for the medium-scale steady state VAR model\label{subsec:Results-for-medium-scale}}

\begin{table}[H]
\begin{centering}
\begin{tabular}{>{\raggedright}p{2cm}>{\raggedright}p{2cm}rrrr}
\hline 
 &  & Standard & BO-EI & BOOP-EI & Grid\tabularnewline
\cline{3-6} \cline{4-6} \cline{5-6} \cline{6-6} 
\multicolumn{2}{l}{Log ML} & $-3078.54$ & $-3052.13$ & $-3052.06$ & $-3052.08$\tabularnewline
\multicolumn{2}{l}{Gibbs iterations} &  & $2.75\cdot10{}^{6}$ & $443660$ & $10^{9}$\tabularnewline
\multicolumn{2}{l}{CPU-time (minutes)} &  & $74$ & $56$ & \tabularnewline
\multicolumn{2}{l}{Model evaluations} &  & $250$ & $250$ & $10^{5}$\tabularnewline
\multicolumn{2}{l}{$\theta_{1}$} & $0.1$ & $0.26$ & $0.27$ & $0.3$\tabularnewline
\multicolumn{2}{l}{$\theta_{2}$} & $0.5$ & $0.37$ & $0.41$ & $0.4$\tabularnewline
\multicolumn{2}{l}{$\theta_{3}$} & $1$ & $0.69$ & $0.76$ & $0.9$\tabularnewline
\hline 
\end{tabular}
\par\end{centering}
\begin{centering}
\caption{Optimization Results Medium Steady-State BVAR.\label{tab:Max}}
\par\end{centering}
{\footnotesize{}The table compares different methods for hyperparameter
optimization in the medium-scale steady-state BVAR. Each method is
run 10 times and the reported hyperparameters for each method are
the best ones over the 10 runs, rounded to two decimals. The marginal
likelihood of the selected models were re-estimated using 200,000
Gibbs iterations with 40,000 as a burn-in. The duration measure is
an average over the 10 runs.}{\footnotesize\par}
\end{table}

Table \ref{tab:Max} summarizes the results from ten runs of the algorithms
for the medium size BVAR model. We see that all three optimization
strategies find hyperparameters that yield substantially higher log
marginal likelihood than the standard values. We can also see that
both Bayesian optimization methods yield as good hyperparameters as
the grid search at only a small fraction of the computational cost.
It is also clear from Table \ref{tab:Max} that a substantial amount
of computations associated with the MCMC are saved when using BOOP.
It is interesting to note that the values for $\theta_{1}\text{ and }\theta_{2}$
are similar for all three optimization approaches but that $\theta_{3}$
differs to some extent. This is due to the flatness of the log marginal
likelihood in that area.

\begin{figure}[H]
\begin{centering}
\includegraphics[scale=0.33]{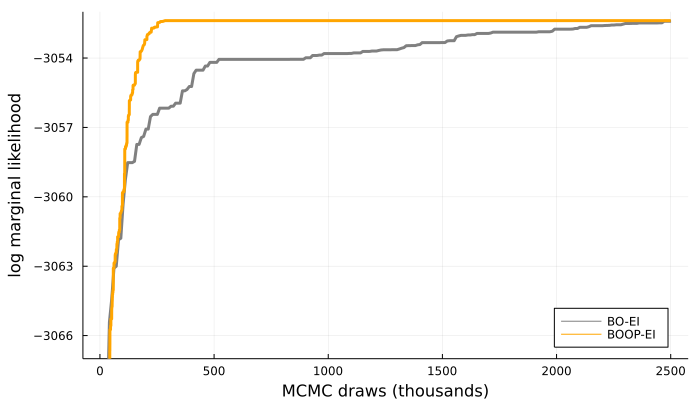}\includegraphics[scale=0.33]{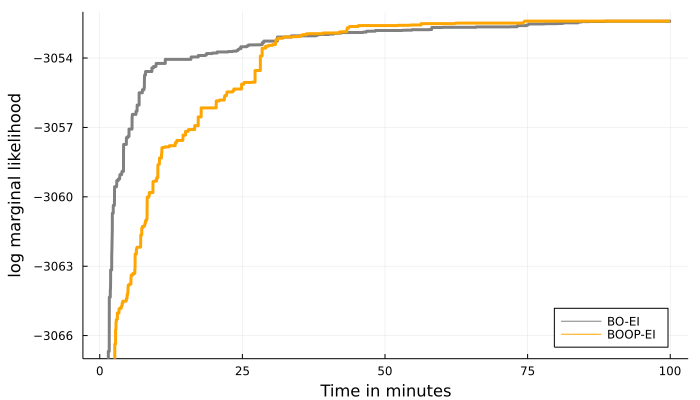}
\par\end{centering}
\caption{Comparison of the convergence speed of the Bayesian optimization methods
as a function of the number of MCMC draws (left) and computing time
(right).\label{fig:Speed-to-Maximum}}
\end{figure}

The left graph of Figure \ref{fig:Speed-to-Maximum} shows that BOOP-EI
finds higher values of the log marginal likelihood using much fewer
MCMC iterations than plain BO with EI acquisitions. From Table \ref{tab:Max}
we can see that BOOP-EI uses, on average, less than a fifth of the
MCMC iterations compared to BO-EI for a full run. Interesting to note
is that BO-EI leads to (on average) a higher number of improvements
on the way to the maximum; while BOOP-EI gives fewer improvements
but of larger magnitude; the strategy of cheaply exploring new territories
before locally optimizing the function pays off. The graph to the
right in Figure \ref{fig:Speed-to-Maximum} shows that for this application
BO-EI is quicker in terms of CPU time to reach fairly high values
for the log marginal likelihood. We see at least two explanations
for this: first, BOOP-EI tries to explore more unknown territories
since they are presumed to be cheap while BO-EI more greedily focuses
on local optimization. Second, and more importantly, the overhead
cost associated with the BOOP-EI acquisition is relatively large in
this medium-sized application where the cost of evaluating the marginal
likelihood itself is not excessive. The fact that BOOP can heavily
reduce the number of MCMC draws while still giving similar CPU computational
time suggest that it is most useful in cases where each log marginal
likelihood evaluation is expensive; this will be demonstrated in the
more computationally demanding models in Sections \ref{subsec:Results-for-the-largescale}
and \ref{sec:Time-varying-parameter-BVAR}.

\begin{figure}[H]
\begin{centering}
\includegraphics[scale=0.36]{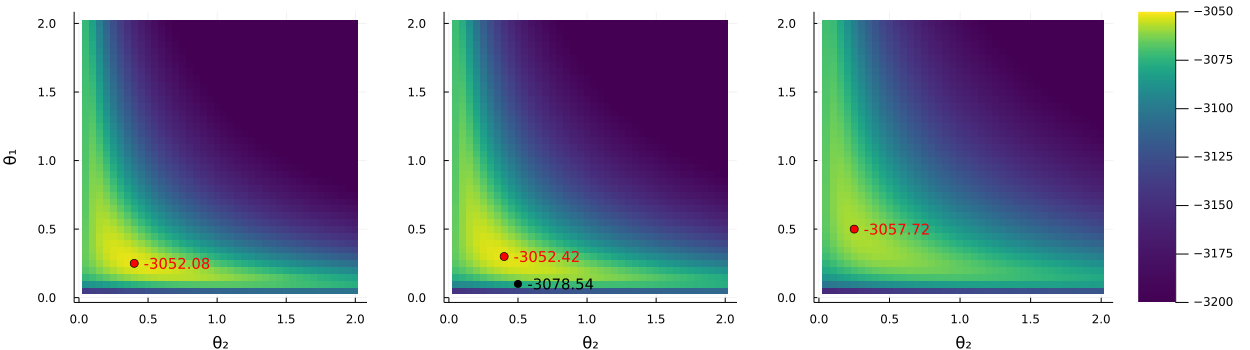}
\par\end{centering}
\caption{Log marginal likelihood surfaces over a fine grid of $(\theta_{1},\theta_{2})$-values.
The hyperparameter values for the lag-decay is $\theta_{3}=0.76$,
(b) $\theta_{3}=1$, (c) $\theta_{3}=2$ (left to right). The red
dot denotes the maximum log marginal likelihood value for the given
$\theta_{3}$ and the black dot, in the middle plot, show the standard
values.\label{fig:Grid}}
\end{figure}

\begin{figure}[H]
\begin{centering}
\includegraphics[scale=0.36]{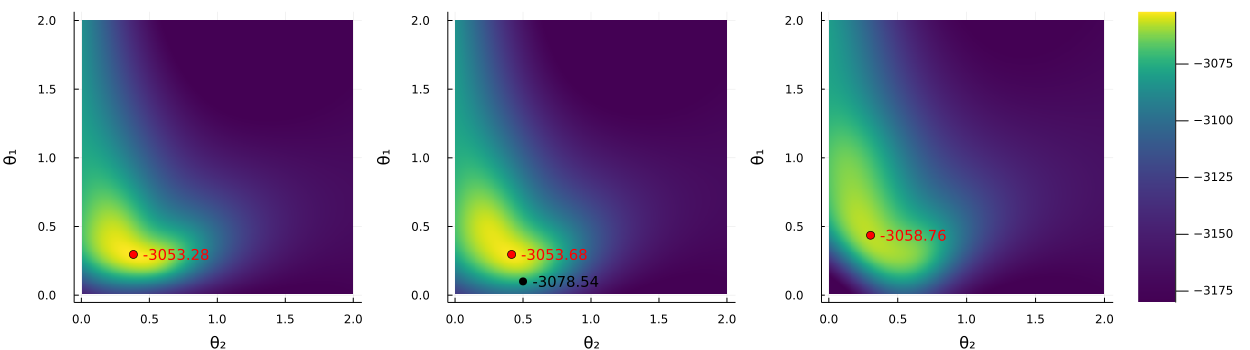}
\par\end{centering}
\caption{GP predictions of the hyperparameter surfaces in Figure \ref{fig:Grid}
based on 250 evaluations for one BOOP-EI run. The hyperparameter for
the lag-decay are $\theta_{3}=0.76,1$, and $2$ (left to right).
Red dot indicates the highest predicted value in the sub-plot and
the black dot, in the middle plot, show the standard values. \label{fig:GPprog-1}}
\end{figure}

Figure \ref{fig:Grid} displays the log marginal likelihood surfaces
over the grid of $(\theta_{1},\theta_{2})$-values used in the grid
search. Each sub-graph is for a fixed value of $\theta_{3}\in\{0.76,1,2\}$.
The red dot indicates the predicted maximum log marginal likelihood
for the given $\theta_{3},$ and the black dot in the middle sub-figure
indicates the standard values. We can see that the standard values
are located outside the high-density region, relatively far away from
the maximum. A comparison of Figures \ref{fig:Grid} and \ref{fig:GPprog-1}
shows that the GP's predicted log marginal likelihood surface is quite
accurate already after merely 250 evaluations; this is quite impressive
considering that Bayesian optimization tries to find the maximum in
the fastest way, and does not aim to have high precision in low-density
regions.

\subsection{Results for the large-scale steady state VAR model\label{subsec:Results-for-the-largescale}}

We also optimize the parameters of the more challenging large BVAR
model containing the 22 different time series, using 250 iterations
for both BO-EI and BOOP-EI. A complete grid search is too costly here,
so we instead compare with parameters obtained from BOOP in the medium-sized
BVAR in Section \ref{subsec:Results-for-medium-scale}, which is a
realistic strategy in practical work.

\begin{table}[H]
\begin{centering}
\begin{tabular}{lrrrr}
\hline 
 & Standard & BO-EI & BOOP-EI & Medium BVAR\tabularnewline
\cline{2-5} \cline{3-5} \cline{4-5} \cline{5-5} 
Log ML & $-7576.31$ & $-7402.50$ & $-7401.09$ & $-7532.61$\tabularnewline
Sd log ML & $0.54$ & $0.81$ & $0.16$ & $0.49$\tabularnewline
Gibbs iterations &  & $3.75\times10^{6}$ & $1.8\times10^{6}$ & \tabularnewline
CPU-time (hours) &  & $64.90$ & $20.22$ & \tabularnewline
$\theta_{1}$ & $0.1$ & $0.47$ & $0.56$ & $0.27$\tabularnewline
$\theta_{2}$ & $0.5$ & $0.06$ & $0.05$ & $0.41$\tabularnewline
$\theta_{3}$ & $1$ & $1.46$ & $1.51$ & $0.76$\tabularnewline
\hline 
\end{tabular}
\par\end{centering}
\caption{\label{tab:BOlarge}Optimization Results Large Steady-State BVAR.}

{\footnotesize{}Hyperparameter optimization in the large-scale steady-state
BVAR. The column named ``Medium BVAR'' are the values obtained from
using BOOP-EI for the medium size model. Both optimization methods
were run 5 times and the reported hyperparameters for each method
are the best ones over the 5 runs, rounded to two decimals. The marginal
likelihood of the selected models were re-estimated using 100,000
Gibbs iterations with 40,000 as a burn-in. The duration measures are
averages over the 5 runs.}{\footnotesize\par}
\end{table}

\begin{figure}[H]
\includegraphics[scale=0.36]{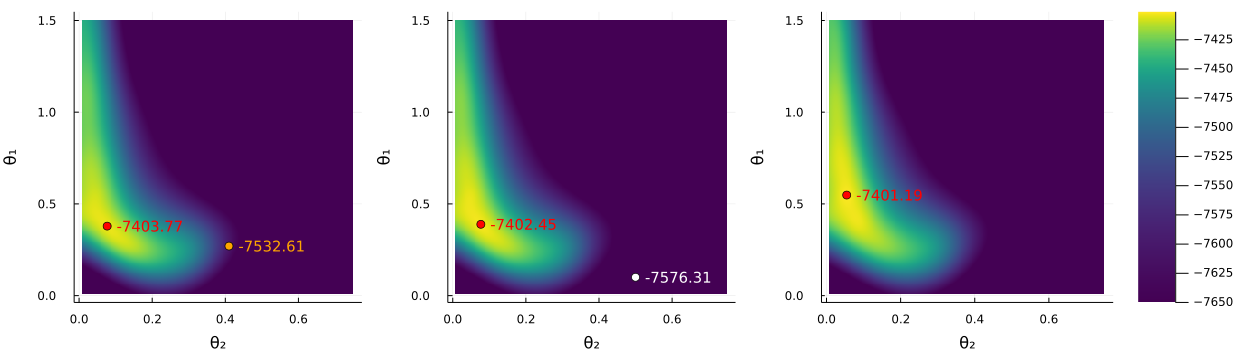}

\caption{GP predictions of the hyperparameter surfaces for the large BVAR based
on 250 evaluations for one BOOP-EI run. The hyperparameter for the
lag-decay are $\theta_{3}=0.76$ (left graph, optimal in medium-size
BVAR), $\theta_{3}=1$ (middle graph, standard value) and $\theta_{3}=1.51$
(right, optimal for BOOP-EI). Red dot indicates the highest predicted
value in the sub-plot. The orange dot in the leftmost plot show the
hyperparameters obtained from BOOP in the medium-sized model and the
white dot in the middle plot show the standard values. \label{fig:GPprogLarge}}
\end{figure}

Table \ref{tab:BOlarge} shows that our method, again, finds optimal
hyperparameters with dramatically larger log ML than standard values,
and also substantially better values than those that are optimal for
the medium-scale BVAR. Finally, note that the hyperparameters selected
by BOOP-EI in the large-scale BVAR are quite different from those
in the medium-scale model. The optimal $\theta_{1}$ applies less
baseline shrinkage than before, but the lag decay ($\theta_{3}$)
is higher, and in particular, the cross-lag shrinkage, $\theta_{2}$,
is much closer to zero, implying much harder shrinkage towards univariate
AR-processes. This latter result strongly suggests that the computationally
attractive conjugate prior structure is a highly sub-optimal solution
since such a prior requires that $\theta_{2}=1$. We can see that
for this more computationally demanding model BOOP-EI is much faster
and finish on average in a third of the time than the regular BO-EI
strategy. Figure \ref{fig:GPprogLarge} show the predicted log marginal
likelihood surface obtained from the last GP in a BOOP run. The rightmost
graph conditions on $\theta_{3}=1.51$, which is optimal for BOOP-EI,
so this graph therefore has the GP with the highest accuracy.

\section{Time-varying parameter BVAR with stochastic volatility\label{sec:Time-varying-parameter-BVAR}}

\subsection{Model and setup}

The time-varying parameter BVAR with stochastic volatility (TVP-SV
BVAR) in \citet{chan2018} is given by 
\begin{equation}
\boldsymbol{B}_{0t}\mathbf{y}_{t}=\boldsymbol{\mu}_{t}+\boldsymbol{B}_{1t}\mathbf{y}_{t-1}+\dots+\boldsymbol{B}_{pt}\mathbf{y}_{t-p}+\boldsymbol{\boldsymbol{\varepsilon}}_{t},\hspace{1em}\boldsymbol{\boldsymbol{\varepsilon}}_{t}\sim N(\boldsymbol{0},\boldsymbol{\Sigma}_{t})
\end{equation}
where $\boldsymbol{\mu}_{t}$ is a vector of time varying intercepts,
$\boldsymbol{B}_{1t},\dots,\boldsymbol{B}_{pt}$ are $n\times n$
matrices of VAR coefficients, $\boldsymbol{B}_{0t}$ is a $n\times n$
lower triangular matrix with ones on the main diagonal. The evolution
of $\boldsymbol{\Sigma}_{t}=\mathrm{diag}\left(\exp(h_{1t}),\dots,\exp(h_{nt})\right)$
is modelled by the vector of log volatilities, $\mathbf{h}_{t}=(h_{1t},\dots,h_{nt})^{\top}$,
evolving as a random walk
\begin{equation}
\mathbf{h}_{t}=\mathbf{h}_{t-1}+\boldsymbol{\zeta}_{t},\qquad\boldsymbol{\zeta}_{t}\overset{\mathrm{iid}}{\sim}N(\mathbf{0},\boldsymbol{\Sigma}_{h}),
\end{equation}
where $\boldsymbol{\Sigma}_{h}=\mathrm{diag}(\sigma_{h_{1}}^{2},\dots,\sigma_{h_{n}}^{2})$
and the starting values in $\mathbf{h}_{0}$ are parameters to be
estimated. Following \citet{chan2018}, we collect all parameters
of $\boldsymbol{\mu}_{t}$ and the $\mathbf{B}_{it}$ matrices in
a $k_{\gamma}$-dimensional vector $\boldsymbol{\gamma}_{t}$ and
write the model in state space form as

\begin{align}
\mathbf{y}_{t} & =\boldsymbol{X}_{t}\boldsymbol{\gamma}_{t}+\boldsymbol{\boldsymbol{\varepsilon}}_{t},\hspace{1em}\boldsymbol{\boldsymbol{\varepsilon}}_{t}\sim N(\boldsymbol{0},\boldsymbol{\Sigma}_{t})\nonumber \\
\boldsymbol{\boldsymbol{\gamma}}_{t} & =\boldsymbol{\boldsymbol{\gamma}}_{t-1}+\boldsymbol{\eta}_{t},\hspace{1em}\boldsymbol{\eta}_{t}\sim N(\boldsymbol{0},\boldsymbol{\Sigma}_{\gamma}),
\end{align}
where $\boldsymbol{X}_{t}$ contain both current and lagged values
of $\mathbf{y}$, $\boldsymbol{\Sigma}_{\gamma}=\mathrm{diag}(\sigma_{\gamma_{1}}^{2},\dots,\sigma_{\gamma_{k\gamma}}^{2})$
and the initial values for the state variables follow $\boldsymbol{\boldsymbol{\gamma}}_{0}\sim N(\mathbf{a}_{\boldsymbol{\gamma}},\mathbf{V}_{\boldsymbol{\gamma}})$
and $\mathbf{h}_{0}\sim N(\mathbf{a}_{\mathbf{h}},\mathbf{V}_{\mathbf{h}})$.

For comparability we choose to use the same prior setup as in \citet{chan2018}
where the variances of the state innovations follow independent inverse-gamma
distributions: $\sigma_{\gamma_{i}}^{2}\sim IG(\nu_{\gamma_{0}},S_{\gamma_{0}})$
if $\gamma_{i}$ is an intercept, $\sigma_{\gamma_{i}}^{2}\sim IG(\nu_{\gamma_{1}},S_{\gamma_{1}})$
if $\gamma_{i}$ is a VAR coefficient and $\sigma_{h}^{2}\sim IG(\nu_{h},S_{h})$
for the innovations to the log variances. Following \citet{chan2018}
we set $\mathbf{a}_{\boldsymbol{\gamma}}=\boldsymbol{0}$, $\mathbf{V}_{\boldsymbol{\gamma}}=10\cdot I_{k_{\gamma}}$,
$\mathbf{a}_{\mathbf{h}}=0$, $\mathbf{V}_{\mathbf{h}}=10\cdot I_{n}$,
and $\nu_{\gamma_{0}}=\nu_{\gamma_{1}}=\nu_{h}=5$, but we use Bayesian
optimization to find the optimal values for the three key hyperparameters
$S_{\gamma_{0}}$, $S_{\gamma_{1}}$ and $S_{h}$ which controls the
degree of time-variation in the states. We collect the three optimized
hyperparameter in the vector $\boldsymbol{\theta}=(\theta_{1},\theta_{2},\theta_{3})^{\top}$,
where $\theta_{1}=S_{\gamma_{0}}$, $\theta_{2}=S_{\gamma_{1}}$ and
$\theta_{3}=S_{h}$. We optimize over the domain $\{\boldsymbol{\theta}:0\leq\theta_{1}\leq5,0\leq\theta_{2}\leq1,0\leq\theta_{3}\leq5\}$,
which allows for all cases from no time variation in any parameter
to high time variation in all the model parameters.

To estimate the marginal likelihood, \citet{chan2018} first obtain
posterior draws of $\boldsymbol{\gamma},\boldsymbol{h},\Sigma_{\boldsymbol{\gamma}},\Sigma_{\mathbf{h}},\boldsymbol{\gamma}_{0},\mathbf{h}_{0}$
using Gibbs sampling, which are then used to design efficient importance
sampling proposals. The marginal likelihood is

\begin{equation}
p(\mathbf{y})=\mathbf{\boldsymbol{\int}}p(\mathbf{y}|\boldsymbol{\gamma},\mathbf{h},\boldsymbol{\psi})p(\boldsymbol{\gamma}|\boldsymbol{\psi})p(\mathbf{h}|\boldsymbol{\psi})p(\boldsymbol{\psi})d\boldsymbol{\gamma}d\mathbf{h}d\boldsymbol{\psi},
\end{equation}
where $\boldsymbol{\psi}$ collects all the static parameters in $\Sigma_{\boldsymbol{\gamma}},\Sigma_{\mathbf{h}},\boldsymbol{\gamma}_{0},\mathbf{h}_{0}$.
The inner integral w.r.t. $\boldsymbol{\gamma}$ can be solved analytically
and afterwards, we can integrate out $\mathbf{h}$ using importance
sampling to obtain an estimate of the \emph{integrated likelihood}

\begin{equation}
p(\mathbf{y}|\boldsymbol{\psi})=\mathbf{\boldsymbol{\int}}p(\mathbf{y}|\boldsymbol{\gamma},\mathbf{h},\boldsymbol{\psi})p(\boldsymbol{\gamma}|\boldsymbol{\psi})p(\mathbf{h}|\boldsymbol{\psi})d\boldsymbol{\gamma}d\mathbf{h}.
\end{equation}
 The last step is to integrate out the fixed parameters from the integrated
likelihood

\begin{equation}
p(\mathbf{y})=\mathbf{\boldsymbol{\int}}p(\mathbf{y}|\boldsymbol{\psi})p(\boldsymbol{\psi})d\boldsymbol{\psi},
\end{equation}
 which is done using another importance sampler. The two nested importance
samplers put the algorithm in the framework of importance sampling
squared ($IS^{2}$, \citet{tran2013importance}). The Chan-Eisenstat
algorithm is elegantly designed, but necessarily computationally expensive
with a single estimate of the marginal likelihood taking 205 minutes
in MATLAB on a standard desktop \citep{chan2018}. We call their MATLAB
code from Julia using the MATLAB.jl package, illustrating that BOOP
can plug in any marginal likelihood estimator. However, we found that
the standard errors in \citet{chan2018} can be more robustly estimated
using the bootstrap and we have done so here. The cost of bootstrapping
the standard errors only has to be taken once for every Bayesian optimization
iteration, and this cost is negligible compared to the computation
of the log marginal likelihood estimate.

We use quarterly data for the GDP-deflator, real GDP, and the short-term
interest rate in the USA from 1954Q3 to 2014Q4 from \citet{chan2018}
for comparability. In addition, we make use of their Matlab code (with
minor adjustments) for computing the marginal likelihood. This shows
another strength with the BOOP approach, that it works on top of existing
code.

We fix the number of Gibbs sampling iterations and the burn-in period
to $20000$ and $5000$ respectively for both BO-EI and BOOP-EI in
all evaluation points. This simplifies the implementation and does
not make a practical difference since the main part of the computational
cost is spent on the log marginal likelihood estimate from importance
sampling. For BO-EI we use $5000$ log marginal likelihood evaluations
in each new evaluation point, while BOOP-EI starts with $1000$ importance
sampling draw and then takes batches of size $100$ until a maximum
of $5000$ samples has been reached. The initial 1000 draws were enough
to make the estimator approximately unbiased.

\subsection{Results for the TVP-SV BVAR}

Table \ref{tab:Result-for-TVP-BVAR} shows the optimized log marginal
likelihood from three independent runs of BO-EI and BOOP-EI; the hyperparameter
values used in \citet{chan2018} are shown for reference. As expected
both BO and BOOP find better hyperparameters than the ones in \citet{chan2018};
this is particularly true for BOOP which gives an increase in the
marginal likelihood of more than 10 units on the log scale on average.
Interestingly, both BO and BOOP suggest that the stochastic volatilities
should be allowed to move more freely than in \citet{chan2018}, but
that there should be less time variation in the intercepts and VAR
coefficients. This points in the same direction as the results in
\citet{chan2018} who find that shutting down the time variation in
the intercept and VAR coefficients actually increases the marginal
likelihood. Our results indicate that when carefully selecting the
shrinkage parameters by optimization, the VAR-dynamics should in fact
be allowed to evolve over time, but at a slower pace.

\begin{figure}[H]
\includegraphics[scale=0.36]{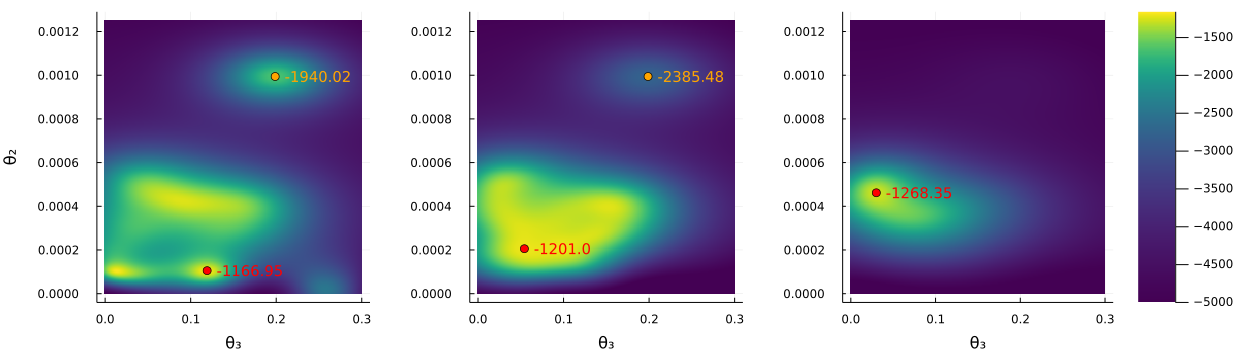}

\caption{Predicted log marginal likelihood over the hyperparameters for stochastic
volatility and the VAR dynamics for $\theta_{1}=0.0086$ (left), 0.05
(middle) and $0.1$ (right). The mode in each plot is marked out by
a red point. A distant local optimum is also marked out by an orange
point. \label{heatStochVol}}
\end{figure}

Table \ref{tab:Result-for-TVP-BVAR} shows a great deal of variability
between runs, in particular for BO. Figure \ref{heatStochVol} shows
that this is probably because the hyperparameter surface is substantially
more complicated and multimodal than for the steady-state BVAR. We
can also see that the log marginal likelihood is relatively insensitive
to changes in $\theta_{3}$ around the mode while it is very sensitive
to changes in $\theta_{2}$.

\begin{center}
\begin{table}[H]
\begin{tabular}{lrrrrrrr}
\hline 
 & CE & BO1 & BO2 & BO3 & BOOP1 & BOOP2 & BOOP3\tabularnewline
\cline{2-8} \cline{3-8} \cline{4-8} \cline{5-8} \cline{6-8} \cline{7-8} \cline{8-8} 
log ML & $-1180.2$ & $-1169.25$ & $-1170.57$ & $-1178.34$ & $-1167.32$ & $-1172.92$ & $-1168.49$\tabularnewline
SE & $0.12$ & $0.89$ & $0.49$ & $0.32$ & $1.24$ & $0.47$ & $1.60$\tabularnewline
$\theta_{1}\times10^{3}$ & $40$ & $19.05$ & $8.66$ & $29.53$ & $7.65$ & $12.22$ & $15.14$\tabularnewline
$\theta_{2}\times10^{5}$ & $40$ & $9.81$ & $10.65$ & $11.07$ & $10.26$ & $7.06$ & $8.70$\tabularnewline
$\theta_{3}\times10^{3}$ & $40$ & $77.56$ & $119.07$ & $25.04$ & $73.81$ & $25.12$ & $114.42$\tabularnewline
Iterations & - & $67$ & $35$ & $46$ & $81$ & $44$ & $157$\tabularnewline
CPU time (hours) & - & $83.40$ & $42.47$ & $56.25$ & $34.90$ & $22.49$ & $77.89$\tabularnewline
\hline 
\end{tabular}\caption{Result for TVP-BVAR with stochastic volatility for three independent
runs of BO and BOOP. CE is taken from Table 3 in \citet{chan2018}.
The row named SE shows the numerical standard errors. Runs were stopped
in case there was no improvement in the last 50 hours.\label{tab:Result-for-TVP-BVAR}}
\end{table}
\par\end{center}

\section{Concluding remarks\label{sec:Conclusion}}

We propose a new Bayesian optimization method for finding optimal
hyperparameters in econometric models. The method can be used to optimize
any noisy function where the precision is under the control of the
user. We focus on the common situation of maximizing a marginal likelihood
evaluated by MCMC or importance sampling, where the precision is determined
by the number of MCMC or importance sampling draws. The ability to
choose the precision makes it possible for the algorithm to take occasional
cheap and noisy evaluations to explore the marginal likelihood surface,
thereby finding the optimum faster.

We assess the performance of the new algorithm by optimizing the prior
hyperparameters in the extensively used BVAR with stochastic volatility
and time-varying parameters and the steady-state BVAR model in both
a medium-sized and a large-scale VAR. The method is shown to be practical
and competitive to other approaches in that it finds the optimum using
a substantially smaller computational budget, and has the potential
of being part of the standard toolkit for BVARs. We have focused on
optimizing the marginal likelihood, but the method is directly applicable
to other score functions, e.g. the popular log predictive score (\citealp{geweke2007smoothly}
and \citealp{villani2012generalized}).

Our approach builds on the assumption that the noisy estimates of
the log marginal likelihoods are approximately unbiased, which we
verify is a reasonable assumption in the three applications if the
first BOOP evaluation is based on a marginal likelihood estimator
from enough MCMC draws. The unbiasedness of the log marginal likelihood
will, however, depend on the combination of MCMC sampler and marginal
likelihood estimator, see \citet{adolfson2007bayesian} for some evidence
from Dynamic Stochastic General Equilibrium (DSGE) models \citep{an2007bayesian}.
For example, the simulations in \citet{adolfson2007bayesian} suggest
that sampling with the independence Metropolis-Hastings combined with
the \citet{chib2001marginal} estimator is nearly unbiased, whereas
sampling with the random walk Metropolis algorithm combined with the
modified harmonic estimator \citep{geweke1999using} can be severely
biased, unless the posterior sample is extremely large. It would therefore
be interesting to extend the method to cases with biased evaluations
where the marginal likelihood estimates are persistent and only slowly
approaching the true marginal likelihood. Since the marginal likelihood
trajectory over MCMC iterations is rather smooth \citep{adolfson2007bayesian}
one can try to predict its evolution and then correct the bias in
the marginal likelihood estimates.

\bibliographystyle{apalike}
\bibliography{ref}

\end{document}